\documentclass[prb,aps,twocolumn,amsmath,amssymb,superscriptaddress,longbibliography]{revtex4-2}
\usepackage[dvips]{epsfig}
\usepackage{float}
\usepackage{color,soul}
\usepackage[sort&compress]{natbib}
\usepackage[english]{babel}
\makeatletter
\renewcommand{\@biblabel}[1]{#1. }
\renewcommand{\@dotsep}{500}
\renewcommand{\@pnumwidth}{0em}
\renewcommand{\l@figure}[2]{
\@dottedtocline{1}{1.5em}{2em}{Figure #1}{}\vspace{15pt}}
\makeatletter
\def\bbl@set@language#1{%
  \edef\languagename{%
    \ifnum\escapechar=\expandafter`\string#1\@empty
    \else\string#1\@empty\fi}%
  \@ifundefined{babel@language@alias@\languagename}{}{%
    \edef\languagename{\@nameuse{babel@language@alias@\languagename}}%
  }%
  \select@language{\languagename}%
  \expandafter\ifx\csname date\languagename\endcsname\relax\else
    \if@filesw
      \protected@write\@auxout{}{\string\select@language{\languagename}}%
      \bbl@for\bbl@tempa\BabelContentsFiles{%
        \addtocontents{\bbl@tempa}{\xstring\select@language{\languagename}}}%
      \bbl@usehooks{write}{}%
    \fi
  \fi}
\newcommand{\DeclareLanguageAlias}[2]{%
  \global\@namedef{babel@language@alias@#1}{#2}%
}
\makeatother

\DeclareLanguageAlias{en}{english}
\begin{document}
\title{High-performance Hybrid Lithium Niobate Electro-optic Modulators Integrated with Low-loss Silicon Nitride Waveguides on a Wafer-scale Silicon Photonics Platform}
\author{Arif Rahman}
\affiliation{University of California, San Diego, Department of Electrical and Computer Engineering, La~Jolla, California 92093-0407, USA}
\author{Forrest Valdez}
\affiliation{LIGENTEC SA, EPFL Innovation Park, Batiment L, Ch. de la Dent d'Oche 1B, 1024 Ecublens VD, Switzerland}
\author{Viphretuo Mere}
\affiliation{LIGENTEC SA, EPFL Innovation Park, Batiment L, Ch. de la Dent d'Oche 1B, 1024 Ecublens VD, Switzerland}
\author{Camiel Op de Beeck}
\email{cob@ligentec.com}
\affiliation{LIGENTEC SA, EPFL Innovation Park, Batiment L, Ch. de la Dent d'Oche 1B, 1024 Ecublens VD, Switzerland}
\author{Pieter Wuytens}
\affiliation{LIGENTEC SA, EPFL Innovation Park, Batiment L, Ch. de la Dent d'Oche 1B, 1024 Ecublens VD, Switzerland}
\author{Shayan Mookherjea}
\email{smookherjea@ucsd.edu}
\affiliation{University of California, San Diego, Department of Electrical and Computer Engineering, La~Jolla, California 92093-0407, USA}
\date{April 3, 2025}
\begin{abstract}	
Heterogeneously-integrated electro-optic modulators (EOM) are demonstrated using the hybrid-mode concept, incorporating thin-film lithium niobate (LN) by bonding with silicon nitride (SiN) passive photonics. At wavelengths near 1550 nm, these EOMs demonstrated greater than 30 dB extinction ratio, 3.8 dB on-chip insertion loss, a low-frequency half-wave voltage-length product ($V_\pi L$) of 3.8 V.cm, and a 3-dB EO modulation bandwidth exceeding 110 GHz. This work demonstrates the combination of multi-layer low-loss SiN waveguides with high-performance LN EOMs made in a scalable fabrication process using conventional low-resistivity silicon (Si) wafers.  
\end{abstract}
\maketitle
%-----------------------------------------------------------------------
%-----------------------------------------------------------------------
\section{Introduction}
\label{sec-introduction}
The integration of high-performance electro-optic modulator (EOM) devices with a low-loss photonic integrated circuit (PIC) platform can address important challenges in modern photonics, such as developing high-throughput interconnects and switches, signal processing applications including wideband waveform generation and low-noise signal acquisition, and advanced computing. Traditionally, the most widely-used EOM devices are made using lithium niobate (LN) which supports a wide range of optical wavelengths and radio frequencies (RF)~\cite{wootenReviewLithiumNiobate2000,howertonBroadbandTravelingWave2002}. However, traditional LN EOM devices require specialized fabrication methods and materials which are different from those used in the global silicon (Si) microelectronics industry. On the other hand, Si photonic EOM devices, which are made in Si foundries, have shown limited performance with lower modulation amplitude at higher speeds, and limited linearity and optical power handling due to free-carrier absorption and two-photon absorption~\cite{yuTradeoffOpticalModulation2014,witzensHighSpeedSiliconPhotonics2018,jainHighSpurFreeDynamic2019,shekharRoadmappingNextGeneration2024}. Recent interest in thin-film lithium niobate (TFLN) or lithium niobate on insulator (LNOI) photonics~\cite{poberajLithiumNiobateInsulator2012} has led to very high bandwidth modulators using a variety of designs and fabrication approaches~\cite{wangIntegratedLithiumNiobate2018,weigelBondedThinFilm2018a,mercanteThinFilmLithium2018,heHighperformanceHybridSilicon2019}. An important challenge now is to make high-performance LN EOM devices using industry-standard Si wafer fabrication methods, which could lead to greater scalability and lower cost, and the ability to make more complex designs with higher yield. 

One approach to developing such modulators is to use the hybrid-mode design, in which the light is distributed among two (or more) waveguide core materials, one of which is a thin-film slab of LN, and the other can be a dielectric or semiconductor (e.g., silicon nitride, SiN, or Si) which is patterned into strip waveguides~\cite{chenHybridSiliconLithium2014,weigelLightwaveCircuitsLithium2016a,weigelBondedThinFilm2018a,boyntonHeterogeneouslyIntegratedSilicon2020}. 
This design can be contrasted to one in which the LN slab was loaded by a dielectric material to form a rib waveguide~\cite{leeHybridSiLiNbO3Microring2011,rabieiHeterogeneousLithiumNiobate2013,bakishVoltageInducedPhaseShift2013}. In the hybrid-mode devices such as those discussed here, the waveguide core materials were separated by a thin cladding layer, such as silicon dioxide, and the strip waveguides in the dielectric material were used for passive photonic components without LN. The thin LN layer was transferred by bonding to a planarized wafer containing the strip waveguides~\cite{mereModularFabricationProcess2022,mereImprovedFabricationScalable2023}. After bonding, the LN slab can be left unetched because the mode confinement in the EO region can be controlled over a wide range by changing only the width of the strip waveguide~\cite{weigelLightwaveCircuitsLithium2016a}. Wide waveguides, which confine most of the light in the passive material, are used in the feeder section and for waveguide bends, splitters and combiners. Narrow waveguides, which push light mostly into the LN region, are used in the phase-shift section of the EOM device~\cite{weigelLightwaveCircuitsLithium2016a,weigelBondedThinFilm2018a}. As design and fabrication technology have improved, the mode confinement factor in LN in the modulation section of hybrid LN EOM devices has increased from 11\%~\cite{chenHybridSiliconLithium2014} to over 80\%~\cite{weigelBondedThinFilm2018a}.  

The hybrid-mode design does not require precise alignment during film transfer, in contrast to micro-transfer printing~\cite{smithHybridIntegrationChipscale2022,vandekerckhoveReliableMicrotransferPrinting2023}, and does not require a smooth and damage-free LN waveguide etching or milling process, in contrast to EOM devices developed on an LNOI platform~\cite{guarinoElectroOpticallyTunable2007,wangIntegratedLithiumNiobate2018,heHighperformanceHybridSilicon2019,chenHighPerformanceThinfilm2022,dellatorreFoldedElectroopticalModulators2025}. Furthermore, heterogeneous integration of thin-film EO crystals is a powerful tool for incorporating high-performance EO microsystems by building upon existing Si and SiN photonics technology without having to start from scratch, since many components of a mature process design kit (PDK) can be reused in the design of large-scale PICs~\cite{mookherjeaThinfilmLithiumNiobate2023}. Because of these attractive characteristics, here we use the hybrid-mode concept to integrate high-performance EOM devices with a low-loss SiN photonics platform. 

%-----------------------------------------------------------
\begin{figure*}[tbh]
\centering
\includegraphics[width=\linewidth, clip=true]{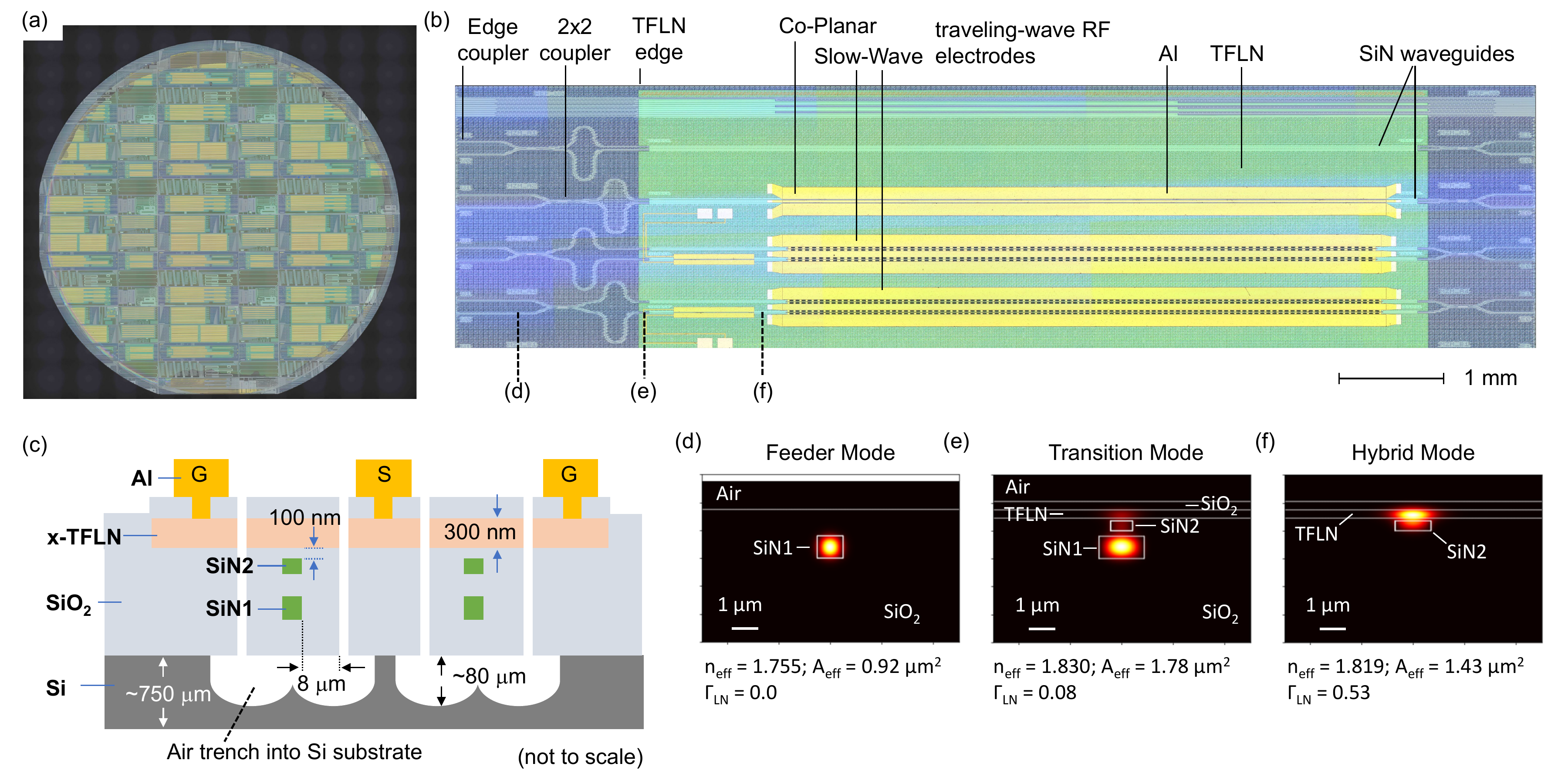}
\caption{(a) Photograph of the fabricated wafer (100~mm diameter). (b) Magnified photograph of the electro-optic modulator (EOM) devices discussed in this report. (c) Schematic drawing of the wafer cross-section. Abbreviations: S and G: signal and ground electrodes, respectively; TFLN: thin-film lithium niobate; SiN: silicon nitride; Si: silicon; Al: aluminum. (d)-(f) Simulations of the optical mode in the feeder, transition, and hybrid sections, which are indicated in panel (b). The optical effective index ($n_\text{eff}$), effective area ($A_\text{eff}$) and mode confinement fraction in LN ($\Gamma_\text{LN}$) are tabulated.} 
\label{fig-wafer}
\end{figure*}
%-----------------------------------------------------------

%-----------------------------------------------------------------------
\section{Description}
\label{sec-description}
A photograph of the completed wafer is shown in Fig.~\ref{fig-wafer}(a) and the three EOM devices discussed in this report are shown in Fig.~\ref{fig-wafer}(b). This image also shows some passive structures (without electrodes) that were included for measurement calibration. These EOM devices are push-pull Mach-Zehnder interferometers (MZI) in which both the optical and microwave fields co-propagate from left to right. A portion of the bonded LN film (x-cut orientation, 300 nm thickness) which covers the SiN waveguides is visible in the figure with a slightly different color. Figure~\ref{fig-wafer}(c) shows a schematic drawing of the wafer cross-section with two silicon nitride layers (SiN1 and SiN2) with strip waveguides. The SiN1 layer contains the feeder waveguides and $2 \times 2$ multi-mode interference (MMI) couplers (length about $100\ \mathrm{\mu m}$) and an optical path length difference segment in the passive region outside the LN (thus creating an asymmetric MZI structure). The feeder waveguides continue a short distance into the LN-covered region. An adiabatic taper was formed between the SiN1-SiN2 layers by decreasing the width of the SiN1 waveguide in the lower layer while simultaneously increasing the width of the SiN2 waveguide in the upper layer. The SiN2 waveguide tapers from the transition mode to the hybrid mode formed between the SiN2 strip waveguide and the LN slab, which is used in the EO phase-shift segments.  

The optical and microwave structures reported here were designed using software (Ansys Lumerical; and PathWave EM Design, Keysight Technologies) guided by insights from earlier work~\cite{wangAchievingBeyond100GHzLargesignal2019a,valdez110GHz1102022a,valdezIntegratedCbandSiliconlithium2023,valdez100GHzBandwidth2023}. As shown in Fig.~\ref{fig-wafer}(d)-(f), the SiN1 waveguide has the smallest optical effective area ($A_\text{eff} = 0.92\ \mathrm{\mu m}^2$), whereas a representative mode in the transition section and the hybrid mode have a larger effective area of $1.78\ \mathrm{\mu m}^2$ and $1.43\ \mathrm{\mu m}^2$, respectively. The confinement factor $\Gamma$ stated in Figs.~\ref{fig-wafer}(d)-(f) is defined by the spatial distribution of the magnitude of the Poynting vector, $\mathbf{S} = \mathbf{E}\times \mathbf{H}^*$ where $\mathbf{E}$ and $\mathbf{H}$ are the vectorial electric and magnetic field distributions of the mode. The hybrid mode has $\Gamma_\mathrm{LN} \approx 53\%$ of the optical power in LN.  

The three EOM designs shown in Fig.~\ref{fig-wafer}(b) from top to bottom are labeled Designs 1, 2, and 3, respectively. Design 1 uses coplanar waveguide (CPW) electrodes with a signal-ground gap distance of $G=6\ \mathrm{\mu m}$. According to our simulations, Design 1 can achieve a good match between the RF phase and optical group indices of refraction up to high RF modulation bandwidths. However, the SiN waveguides were fabricated on a low-resistivity Si substrate, for which the high-frequency RF propagation loss is too large to support high modulation bandwidths (as will be shown by the data presented in this report). Therefore, in Designs 2 and 3, portions of the substrate were removed by etching air trenches as described below (also see the Supplementary Information, Section S1) whereas no trenches were included with Design 1, in order to provide a comparison. Removing the low-resistivity Si substrate reduces the propagation loss of the RF coplanar waveguide mode and increases the modulation bandwidth~\cite{xueBreakingBandwidthLimit2022,panCompactSubstrateremovedThinfilm2022,xuCMOSlevelvoltageSubstrateremovedThinfilm2022,chenHighPerformanceThinfilm2022}. These trenches also have the effect of lowering the RF index. To re-establish index matching in Designs 2 and 3, we used periodically-patterned slow-wave electrodes (SWE)~\cite{sakamotoNarrowGapCoplanar1995}. Specifically, Designs 2 and 3 used $G=4\ \mathrm{\mu m}$ and $5\ \mathrm{\mu m}$, respectively, with a capacitively-loaded T-rail electrode design~\cite{rosaMicrowaveIndexEngineering2018,valdezIntegratedCbandSiliconlithium2023}. As the distance between the inner electrode edge and the inner T-rail edge increases, the RF wave becomes slower which results in a larger RF effective index. Furthermore, the T-rail stem width also affects the RF wave velocity, with narrower width corresponding to a slower RF field. The measurements reported below show that both Designs 2 and 3 can support high modulation bandwidths beyond 100 GHz. 

%-----------------------------------------------------------------
\section{Fabrication process}
\label{sec-fabrication}
\begin{figure*}[tbh]
\centering
\includegraphics[width=\linewidth, clip=true]{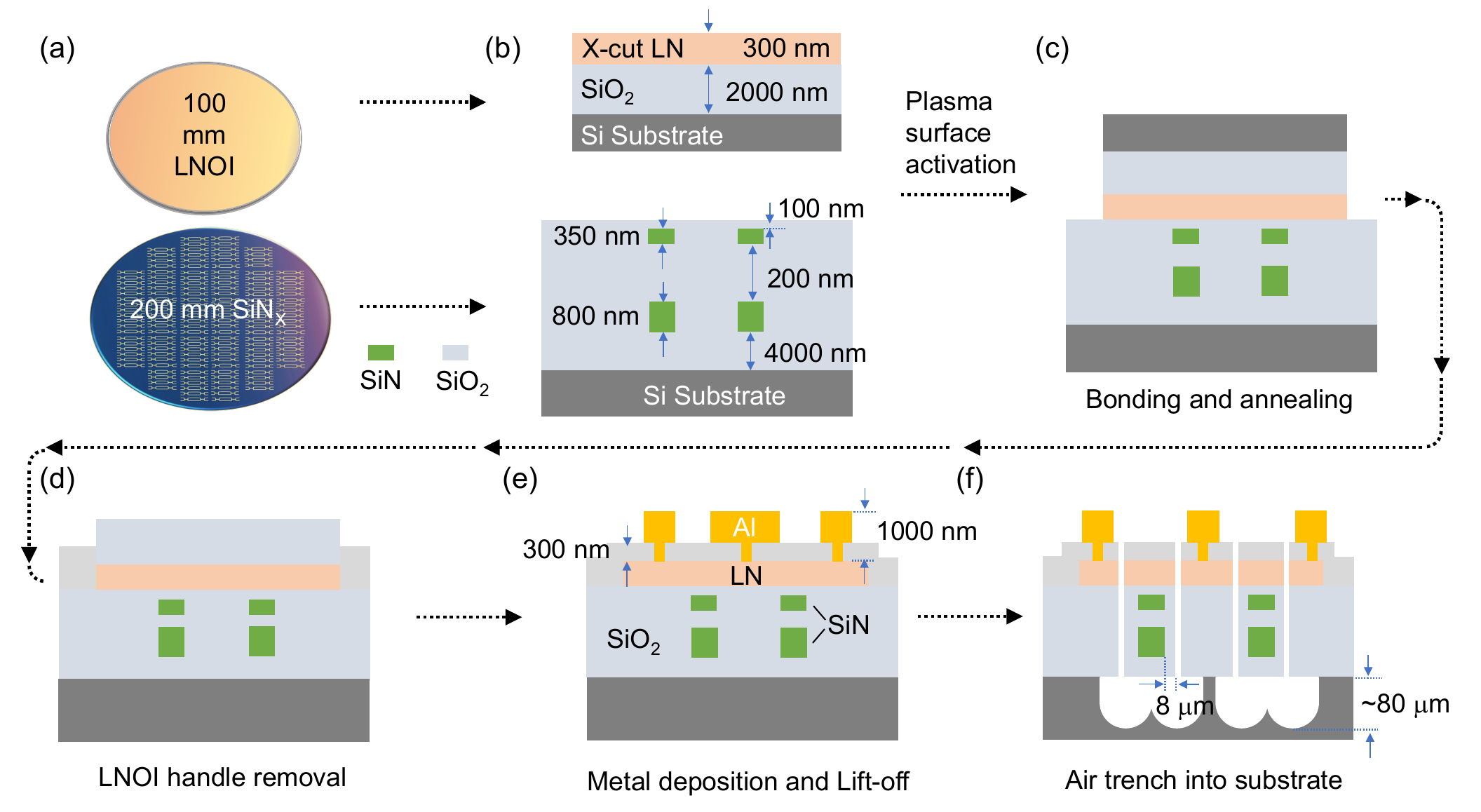}
\caption{Process flow diagram showing the main steps, (a) - (f), in the fabrication of the hybrid TFLN-SiN modulators. (Schematic diagrams, not to scale.)} 
\label{fig-fabrication}
\end{figure*}

Figure~\ref{fig-fabrication} describes the fabrication process, which starts with a 200~mm diameter silicon (Si) handle wafer, with a thickness of about $700\ \mathrm{\mu m}$ and resistivity of about $10 \ \mathrm{\Omega .cm}$. The two SiN layers were fabricated using LIGENTEC's proprietary technology with high-quality silicon nitride deposited using a low-pressure chemical-vapor deposition (LPCVD) process. The strip waveguides in the SiN1 layer have dimensions of $1 \ \mathrm{\mu m} \times 0.8 \ \mathrm{\mu m}$ (width $\times$ thickness), and those in the SiN2 layer have dimensions of $1.4 \ \mathrm{\mu m} \times 0.35 \ \mathrm{\mu m}$. The buffer oxide below SiN1 has a nominal thickness of $4 \ \mathrm{\mu m}$, and the top oxide layer above SiN2 has a nominal thickness of $0.1 \ \mathrm{\mu m}$. A planar bondable surface was achieved using chemical-mechanical polishing (CMP). After processing, the wafer was cored down to a diameter of 100~mm to match the lithium-niobate-on-insulator (LNOI) wafers used here.  

Lithium-Niobate-On-Insulator (LNOI) wafers (100~mm diameter) with 300~nm thickness x-cut LN on 2 $\mathrm{\mu m}$ thick buried oxide, and an approximately 500 $\mathrm{\mu m}$ thick Si handle were commercially procured (NanoLN, Jinan Jingzheng Electronics Co., Ltd.). After cleaning and surface activation [Fig.~\ref{fig-fabrication}(b)], the Si and LNOI wafers were bonded using a hydrophilic process [Fig.~\ref{fig-fabrication}(c)]. The initial contact bonding was mediated by weak van der Waals forces between the exposed surfaces. After room temperature bonding, the bonded sample was annealed using temperature cycles up to 300 $^\circ \mathrm{C}$ under applied pressure. The annealing process converts the hydrogen bonds into covalent bonds at the interface, thereby improving the bond strength~\cite{xuGlassonLiNbO3HeterostructureFormed2018}. No adhesives or polymers were used as the bonding layer. The handle and oxide layer of the LNOI wafer were removed using a combination of dry and wet etching steps [Fig.~\ref{fig-fabrication}(d)]. The Aluminum (Al) electrodes were formed on top of the LN layer using deposition, lithography, and a dry-etch process [Fig.~\ref{fig-fabrication}(e)]. 

After bonding, LN etching was performed across the wafer to remove the LN over the passive sections where it is not needed, such as over the SiN1 components. The SiN1 waveguides continue outside of the bonded region, as shown in Fig.~\ref{fig-wafer}(b), and can be used to connect seamlessly to other components on the same chip. Vertical trenches were formed from the LN surface through the oxide into the Si substrate as access holes for a dry-etch process which formed an undercut. This is shown schematically in Fig.~\ref{fig-fabrication}(f), and the Supplementary Information, Section S1, shows scanning electron microscope (SEM) images of the cross-section. The trenches have lateral dimensions of about $30\ \mathrm{\mu m} \times 8\ \mathrm{\mu m}$ (length $\times$ width) and were about $5\ \mathrm{\mu m}$ in height. They were placed about $8\ \mathrm{\mu m}$ away from the SiN2 waveguide edge and do not affect the optical mode. These trenches were used to access the Si substrate for an isotropic dry etch process which undercut the phase-shift segments in the Designs 2 and 3 EOM devices. Visual inspection showed a lateral undercut width of about $80\ \mathrm{\mu m}$.    

%-----------------------------------------------------------
%-----------------------------------------------------------

%-----------------------------------------------------------------
\section{Results}
\label{sec-results}
Individual dies were diced from the wafer and were tested using an optoelectronic probe station at room temperature. Continuous-wave light was coupled to the waveguides using a tunable semiconductor laser source (81640B; Keysight Technologies) and lensed tapered fibers with a spot size of about $3.5 \ \mathrm{\mu m}$ (Oz Optics). The on-chip optical power was about -3~dBm (0.5~mW). An optical transmission scan over wavelengths from 1520~nm to 1620~nm is reported in the Supplementary Information, Section S2, and shows a high on-off contrast of about~30 dB over more than 100~nm of operational bandwidth.  

\subsection{Optical insertion loss}
The on-chip insertion loss (IL) of the EOM was measured by comparing the optical transmission through the EOM to that through a passive waveguide formed in the SiN1 layer. Both devices use similar waveguide-fiber edge couplers, but the passive waveguide does not include transitions to and from the SiN2 layer, or the sub-components of the EOM, such as the 3-dB input and output couplers and the phase-shift segments. Since the EOM is based on a MZI which includes a path-length difference segment, the values at the peak of the spectra near 1550 nm were taken as representative of its transmission. The measured average on-chip IL (and one standard deviation) for EOM devices across four chips in the three different designs were: 3.1 dB (1.4 dB) for Design 1, 6.6 dB (2.7 dB) for Design 2, and 4.0 dB (0.8 dB). Design 2 has the smallest signal-ground electrode gap ($G$) and therefore, has the highest IL. The observed variation in IL from one chip to another is a consequence of the varying layer thicknesses and fidelity-to-design of devices fabricated on the edges of the wafer compared to those in the center. As this was the first fabrication run for these hybrid modulators, we believe that uniformity issues can be iteratively improved in future fabrication runs. At this time, we do not have sub-component level IL data (i.e., for the inter-layer transitions alone, or the MMI coupler alone) because of the limited space that was available to us on this fabrication run. 

Waveguide-coupled microring resonators with a waveguide width of $1.7\ \mathrm{\mu m}$ were formed in the SiN1 layer with a free spectral range of 150~GHz and measured to have a loaded quality factor at critical coupling of about 1.2 million, from which we infer a propagation loss of about 0.2~$\text{dB}.\text{cm}^{-1}$ at 1550 nm across the whole wafer. No significant change in loss was observed by comparing the transmission of devices before and after LN bonding and processing. By measuring light propagation through waveguide spirals formed in the SiN2 layer (lengths of 10.4~cm, 15~cm, and 27~cm), we infer a propagation loss ranging between 0.3--1.0~$\text{dB}.\text{cm}^{-1}$ at 1560~nm, depending on the location of the waveguide on the wafer. The large range and higher loss in the SiN2 layer can be attributed to damage of this layer during the etching of the LN layer, and material absorption in the top oxide layer. A more selective LN removal process to avoid such damage is under development, which should result in more uniform propagation losses across the wafer, near the lower limit of the data reported here. Taken together, these observations suggest that the passive SiN1 and SiN2 layers have low optical loss and can be combined with high-bandwidth EOM devices on the same platform.

%-----------------------------------------------------------
\begin{figure*}[tbh]
\centering
\includegraphics[width=\linewidth, clip=true]{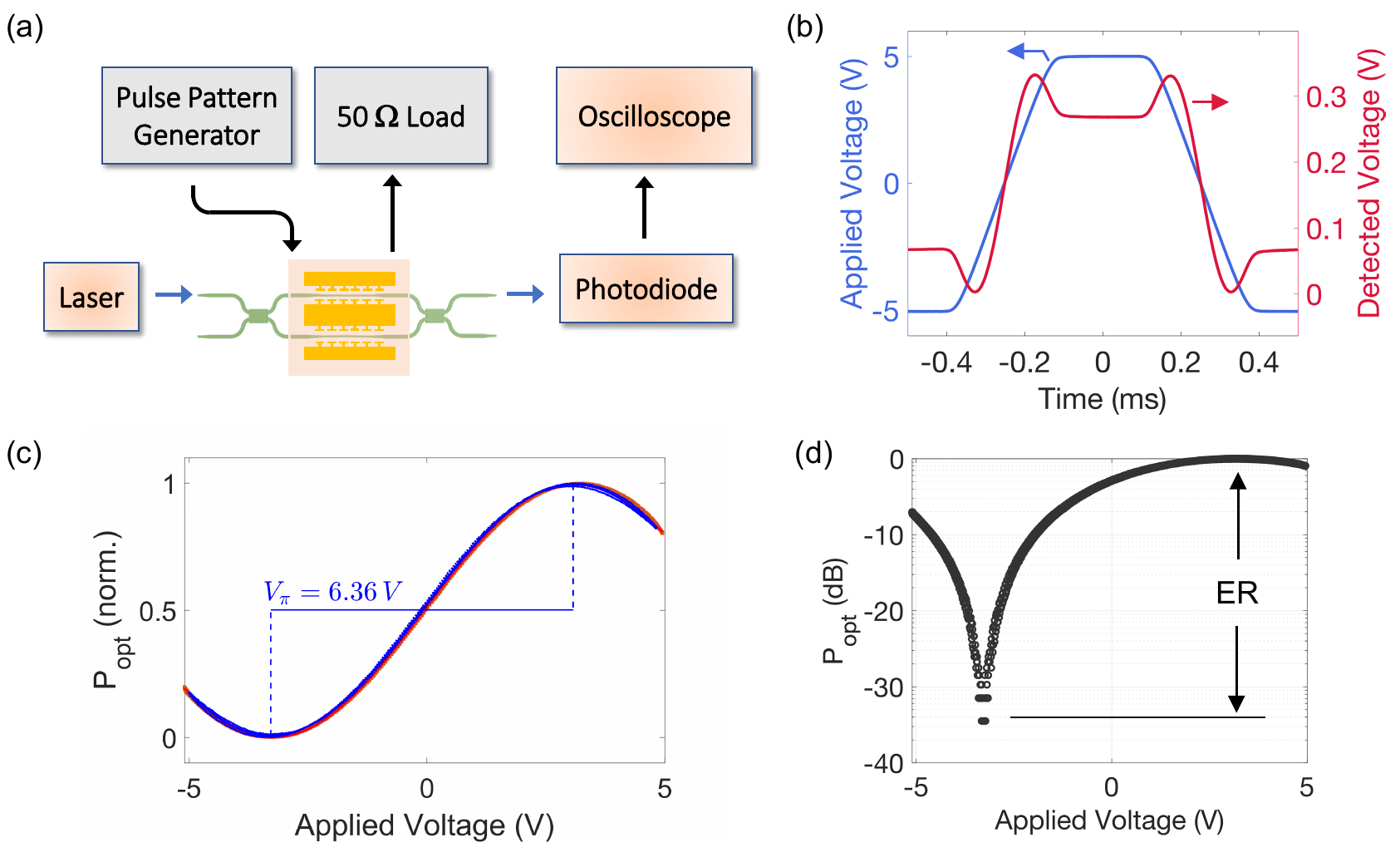}
\caption{(a) Schematic for the measurement of the half-wave voltage, $V_\pi$. (b) Example of the (overdriven) optical response (red line) to the trapezoidal electrical drive signal (blue line). The vertical axis for the optical response is on an arbitrary linear scale determined by the photodetector. (c) Example of the normalized optical power measured as a function of the applied voltage at frequencies of 1~kHz (red) and 1~MHz (blue). The solid line is a fit to the 1 MHz data using the squared-cosine functional form, which yields $V_\pi$ as shown. (d) The same data as in panel (c) for 1~kHz with a logarithmic vertical scale to quantify the extinction ratio, $\mathrm{ER} > 34\ \mathrm{dB}$.} 
\label{fig-VpiL}
\end{figure*}
%-----------------------------------------------------------

\subsection{Half-wave voltage-length product ($\mathrm{V}_\pi L$)}
\label{sec-measure-VpiL}
Figure~\ref{fig-VpiL} describes the half-wave voltage ($\mathrm{V}_\pi$) measurement method and shows the measured data for a representative modulator using Design 3. As shown in the schematic drawing, Fig.~\ref{fig-VpiL}(a), a trapezoidal signal generated by a pulse pattern generator (81110A, Agilent Technologies) was used to drive the modulator with a certain peak-to-peak voltage that exceeds the $\mathrm{V}_\pi$ value anticipated from design simulations. The driving waveform is shown by the blue line in Fig.~\ref{fig-VpiL}(b). The modulated optical signal was detected with a linear photodetector (Model 1817-FC, Newport) and the electrical signal was recorded on an oscilloscope, as shown by the red line in Fig.~\ref{fig-VpiL}(b). Because the applied voltage was greater than $V_\pi$, the over-drive can easily be seen in the oscilloscope trace. This waveform was post-processed using software (MATLAB) to map the optical transmission to the applied voltage, and a cosine-squared fit was used to find $V_\pi$. Trapezoidal waveforms were used because they perform a more stringent test over triangular ramps, as discussed in our earlier work~\cite{valdezIntegratedCbandSiliconlithium2023,mereModularFabricationProcess2022}. Figure~\ref{fig-VpiL}(c) shows the modulated signals as a function of applied voltage for frequencies of 1 kHz (red) and 1 MHz (blue), without a significant difference between the two cases, or for other intermediate frequencies. The data was fit using a cosine-squared equation, resulting in a measured $\mathrm{V}_\pi$ of 6.4 V ($V_\pi L =3.8$ V.cm) in each case. Plotting the 1 kHz data using a logarithmic scale for the vertical axis as shown in Fig.~\ref{fig-VpiL}(d) demonstrates an extinction ratio greater than 34 dB. The average (and one-standard deviation) $\mathrm{V}_\pi L$ values for the three designs measured across four chips were: 4.29 V.cm (0.14 V.cm) for Design 1 [signal-electrode gap $G = 6\ \mathrm{\mu m}$], 3.40 V.cm (0.15 V.cm) for Design 2 [$G = 4\ \mathrm{\mu m}$], and 3.83 V.cm (0.01 V.cm) for Design 3 [$G = 5\ \mathrm{\mu m}$]. As expected, the design with the smallest $G$ has the lowest $\mathrm{V}_\pi L$ value.  

\subsection{3-dB EO modulation bandwidth}
\begin{figure*}[tbh]
\centering
\includegraphics[width=\linewidth, clip=true]{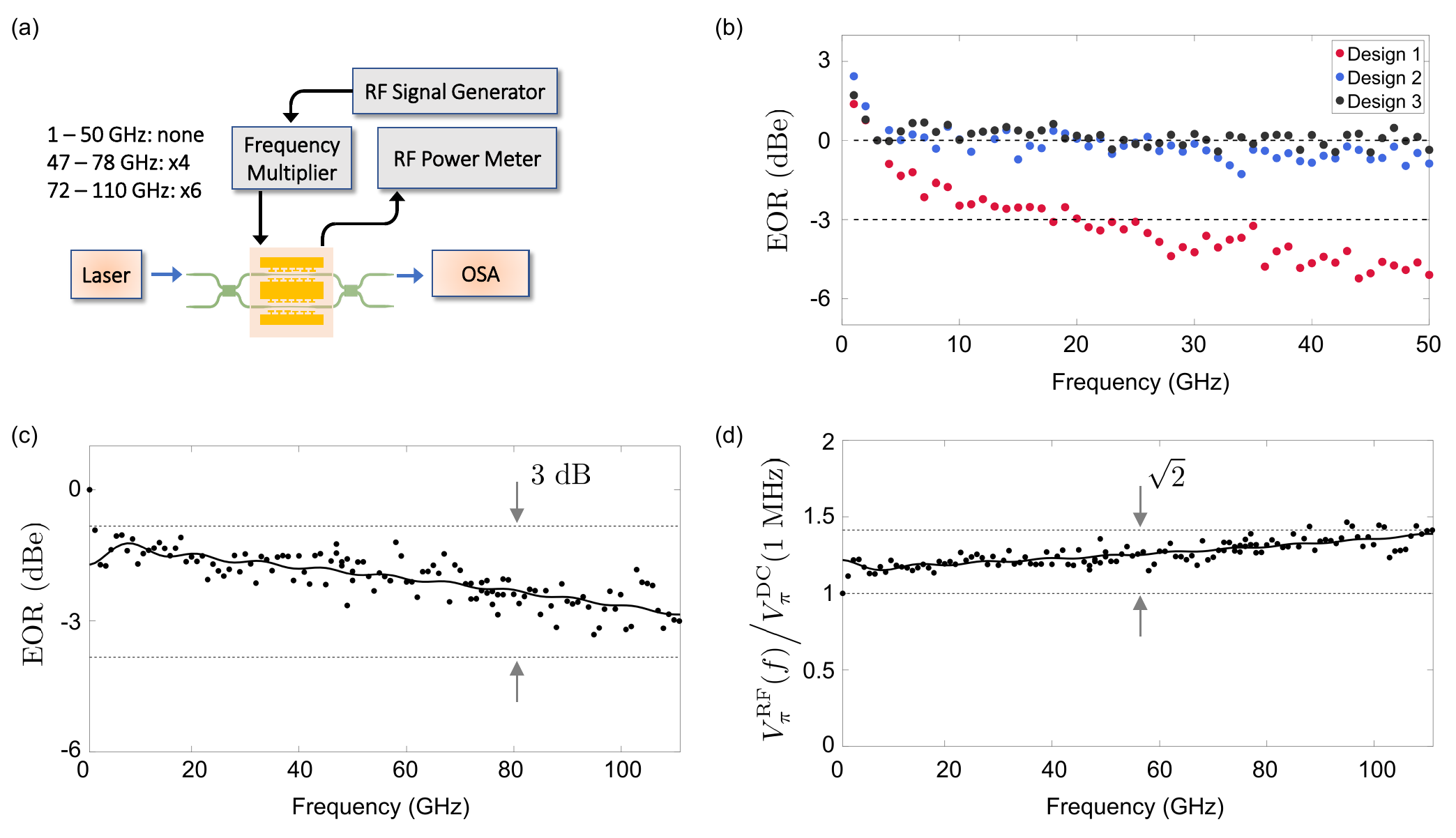}
\caption{(a) Schematic for the measurement of the electro-optic response (EOR). (b) EOR measurement data between 1~GHz and 50~GHz for the three MZM designs on one test chip. (c) EOR measurement data between 1~GHz and 111~GHz for a Design 3 EOM device, normalized to the corresponding EOR value at 1 GHz. The solid black line is a fit to the data using a theoretical model of the EOM. (d) The calculated RF $V_\pi$ as a function of driving frequency using Eq.~\ref{eqn_Vpif}.}
\label{fig-EORdata}
\end{figure*}

As shown schematically in Fig.~\ref{fig-EORdata}(a), we used a modified modulation-sideband method for measuring the electro-optic response (EOR), and our previous work has discussed the similarity of this measurement method to others e.g., based on using a lightwave component analyzer~\cite{valdez110GHz1102022a,valdezIntegratedCbandSiliconlithium2023,valdez100GHzBandwidth2023,valdezBuriedElectrodeHybridBonded2023}. Sinusoidal RF waveforms between 1~GHz and 50~GHz were generated using an RF sweeper (83651B, Hewlett Packard, Inc.). A high-resolution OSA (WaveAnalyzer 1500S, Finisar) was used to measure the spectrum of the modulated signals. High-speed ground-signal-ground RF probes rated to 110~GHz (Model i110, FormFactor) sourced and terminated the traveling-wave electrodes and were used with standard low-loss 1~mm RF cables and connectors. The EOR was determined by tracking the peaks of the carrier signal and the generated sidebands from modulation~\cite{shiHighspeedElectroopticModulator2003,leeBroadbandModulationLight2002,haffnerLowlossPlasmonassistedElectrooptic2018,renaudSub1VoltHighBandwidth2023}.  Active multipliers were used for the 47~GHz to 78~GHz band (SFA-503753420-15SF-E1, Eravant) and for the 72 GHz to 110 GHz band (SFA-753114616-10SF-E1, Eravant). The amount of RF power delivered to the chip was recorded using a calibrated thermal RF power meter (NRP110T.02, Rohde \& Schwarz). The two frequency ranges over which the banded measurements overlap (47--50 GHz and 72--78 GHz) were used to stitch the EOR responses into the composite EOR response. This was accomplished by calculating a mean value for both traces in the overlapping frequency range, and vertically offsetting the second data set to the first curve. Measurements made in the overlapping RF frequency ranges show smooth continuity to within a fraction of a dB. 

Figure~\ref{fig-EORdata}(b) compares the EOR curves for the three EOM designs on one chip. (The Supplementary Information, Section S3, contains similar plots on other test chips, which show similar behavior.) To clearly show the faster drop-off at modulation frequencies above 5~GHz in Design 1 compared to Designs 2 and 3, these EOR curves were each normalized to their value at 3~GHz. (In the literature, the EOR curve for short TFLN modulators measured to high frequencies is usually normalized to its value somewhere in the 1--5 GHz range, because the transmission-line model for the EOR response is usually less valid at low frequencies when the device length is comparable with the RF wavelength.) The main reason for this drop-off in the performance of Design 1 at higher modulation frequencies is the lack of the undercut (trench) of the low-resistivity Si substrate, resulting in higher RF loss and lower EO bandwidth.  

Figure~\ref{fig-EORdata}(c) shows the EOR for the high-bandwidth Design 3 EOM between 1~GHz and 111~GHz, which is the upper limit of our measurement apparatus. In this plot, the EOR is normalized to its value at 1 GHz, which is taken as 0 dBe, similar to our previous work~\cite{valdezIntegratedCbandSiliconlithium2023,valdez100GHzBandwidth2023}. The solid line is a fit to the data using a transmission-line model\cite{ghioneSemiconductorDevicesHighspeed2009} (see Supplementary Information). The EOR of TFLN EOM devices typically show a drop between 1 GHz and 5 GHz before a gradual roll-off is observed over most of the measurement band; this behavior was also shown on different test chips obtained using a lightwave component analyzer (LCA)~\cite{valdezIntegratedCbandSiliconlithium2023}. To examine the EOR roll-off, two horizontal dashed lines were drawn in Fig.~\ref{fig-EORdata}(c) as visual guides. The upper line was drawn at the average EOR value between 1~GHz and 3~GHz, and the lower line was drawn 3 dB below it. The significance of the 3-dB frequency $f_0$ is that the EOR rolls off as $[1+(f/f_0)^2]^{-1}$ as $f$ increases above $f_0$. The EOR response does not roll off below 3 dB up to the measurement limit, and we can therefore take the 3-dB roll-off frequency for these devices as greater than 111 GHz. The Supplementary Information contains similar data on other test chips, showing a 3-dB modulation bandwidth greater than 100~GHz for all tested Design 2 and Design 3 EOMs. 

When considering the overall system requirements to drive an EOM, it is helpful to know $V_\pi$ as a function of the modulation frequencies, because the impact of RF loss, velocity mismatch, and impedance mismatch will also effect the needed voltage for a $\pi$ phase-shift\cite{howertonBroadbandTravelingWave2002}. The driving voltage as a function of modulation frequency $f$ is given by 
%----------------------
\begin{equation} \label{eqn_Vpif}
V_\pi(f) = V_\pi(DC) 10^{-m(f)/20}
\end{equation}
%----------------------
where $V_\pi(DC)$ is the low-frequency half-wave voltage and $m(f)$ is the measured normalized EOR. For a numerical estimate of $V_\pi\mathrm{(DC)}$, we use the half-wave voltage at 1 MHz, measured as described in Section~\ref{sec-measure-VpiL}. Figure~\ref{fig-EORdata}(d) shows the calculated RF $V_\pi(f)$ normalized to $V_\pi\mathrm{(DC)}$. An increase of this ratio by a factor of $\sqrt{2}$ is an alternative definition of the modulation bandwidth because it correlates to the EOR of the modulator (and equivalently, the EO $S_{21}$ value~\cite{valdezIntegratedCbandSiliconlithium2023,pozarMicrowaveEngineering2012}) decreasing by 3 dB (50\%). Based on Fig.~\ref{fig-EORdata}(d), we can also take as the 3-dB modulation bandwidth for this EOM as 111 GHz (the upper limit of our measurement).

%-----------------------------------------------------------
%-----------------------------------------------------------

%-----------------------------------------------------------------
\section{Discussion}
\label{sec-discussion}
Design 1, which uses the CPW waveguides, has the lowest optical insertion loss, but it lacks the substrate trenches and does not support a comparably high modulation bandwidth as the other two designs. Design 2 has a lower $V_\pi L$ product than Design 3, but also has a higher insertion loss. Design 3 offers superior performance across all metrics, with a high extinction ratio, low insertion loss, reasonable $V_\pi L$ product and high 3-dB EO modulation bandwidth. The combination of these attractive performance metrics with a scalable, standard Si wafer based fabrication approach establishes a new state-of-the-art in hybrid LN modulators, in our opinion. Future work can improve performance by studying variations around this benchmark design. In this report, we now include a broader discussion of hybrid modulators as part of a PIC, followed by a comparison with earlier results on hybrid LN modulators also fabricated using a wafer-scale process with either Si or SiN. 

The hybrid optical mode shown in Fig.~\ref{fig-wafer}(f) is suitable for the straight phase-shift sections but does not support tight bends. However, the SiN1 waveguide can be used, for example, to form microring resonators used in frequency comb generation and optical filtering, and for other structures such as folded modulators~\cite{dellatorreFoldedElectroopticalModulators2025}. 

These EOM devices were designed primarily for integration as part of a larger PIC, and not as stand-alone fiber-pigtailed devices. Previously, the fiber-to-waveguide coupling loss for the SiN1 layer waveguides in the LIGENTEC process has been measured to be about 1 dB/facet using lensed fibers with a mode field diameter of $2.5\ \mathrm{\mu m}$. This reference design for the edge couplers was not changed for these EOM devices. Therefore, the potential fiber-to-fiber insertion loss can be estimated as about 2 dB higher than the on-chip IL stated earlier.  

In Ref.~\cite{mereImprovedFabricationScalable2023}, hybrid SiN/LN modulators were demonstrated in which the TFLN layer was not etched after bonding. Here, LN was removed around all waveguides except in the modulator phase-shift sections, so that the established performance of LIGENTEC's passive photonic components in the SiN1 layer was not impacted. In the modulator segments, no etching or patterning of the LN layer was performed for the purposes of optical mode confinement, but some access holes were etched through the LN and oxide layers into the Si substrate for an undercut etch as described earlier. This LN removal process does not require high resolution or low roughness. 

The low-frequency half-wave voltage length product ($V_\pi L$) of a hybrid LN MZM is given by the following equation~\cite{weigelDesignHighbandwidthLowvoltage2021}
\begin{equation}
V_\pi L = \frac{1}{2} \frac{n_\mathrm{eff} \lambda G}{{n_e}^4 r_{33} \Gamma_\mathrm{mo}}
	\label{eq-VpiL}
\end{equation}
where $n_\mathrm{eff}$ is the effective refractive index of the optical mode, $\lambda$ is the wavelength, $G$ is the effective electrode gap distance between the ground and signal lines, $n_\mathrm{e}$ is the extraordinary index of refraction of LN, $r_{33}$ is the linear Pockels coefficient in the crystal z-direction along which the RF fields are oriented, and $\Gamma_\mathrm{mo}$ is the mode overlap integral between the optical mode and RF mode. $L$ is the length of the phase-shift section. The factor of 2 in the denominator is included as the structure is driven in a push-pull configuration. Although Eq.~(\ref{eq-VpiL}) may not perfectly describe SWE structures such as T-rail electrodes, it is a useful indicator of the role of the various physical parameters.

The measured average $V_\pi L$ values of 4.29 V.cm (Design 1), 3.40 V.cm (Design 2) and 3.83 V.cm (Design 3) across the reported test sites can be compared to the values expected from simulations, 2.75 V.cm (Design 1), 3.33 V.cm (Design 2) and 3.96 V.cm (Design 3). Designs 2 and 3 agree quite well with the simulations (difference of about 2\%); however, Design 1 shows a larger difference. Given the limited number of test chips in this batch, we do not have a conclusive explanation for this. Among the many reasons for a difference between the model and the data are the layer thickness and feature size assumptions made in the former, and the site-to-site (field) variations across the fabricated wafer from which the test chips were extracted. The CPW and SWE electrode designs are also differently sensitive to fabrication variations~\cite{wangAchievingBeyond100GHzLargesignal2019a}. Additional clarity into the deviations and statistical uncertainties may be obtained in the future by measuring more devices.  

In previous work on hybrid LN/Si modulators, we have reported $V_\pi L = 2.9$ V.cm for $G = 7 \ \mathrm{\mu m}$ and $V_\pi L = 3.1$ V.cm for $G = 8 \ \mathrm{\mu m}$ and $n_\mathrm{eff} = 2.0$ ~\cite{valdez110GHz1102022}. Here, we achieved $V_\pi L = 3.8$ V.cm for $G = 5 \ \mathrm{\mu m}$ for hybrid SiN modulators with $n_\mathrm{eff} = 1.8$. Using Eq.~(\ref{eq-VpiL}) with these two earlier reference points, the ratio of the mode overlap integrals in the hybrid SiN and Si modulators is $\Gamma_\mathrm{mo}^{\text{SiN}}/\Gamma_\mathrm{mo}^{\text{Si}} = 0.46 - 0.49$. One of the main reasons for the lower value of $\Gamma_\mathrm{mo}$ is that we used a LN thickness of 300~nm here compared to 600~nm in Ref.~\cite{valdez110GHz1102022}. Localizing more of the hybrid mode in LN generally improves the modulation efficiency. Because of the higher refractive index of Si compared to SiN, a thicker LN slab could be used without the hybrid mode becoming too large in cross-sectional area~\cite{wangAchievingBeyond100GHzLargesignal2019a}. Hybrid Si modulators can be inherently more efficient than hybrid SiN modulators, but replacing the Si with SiN has important benefits such as achieving in lower insertion loss and potentially supporting higher optical power levels and a wider optical transparency window. 

Compared to the hybrid LN/SiN modulators reported in Ref.~\cite{churaevHeterogeneouslyIntegratedLithium2023} ($V_\pi L = 8.8$ V.cm with $G = 7 \ \mathrm{\mu m}$ for a single-arm design, from which we can infer $V_\pi L = 4.4$ V.cm for a push-pull modulator), these modulators have an improved efficiency. We show a 3-dB EO bandwidth beyond 100 GHz whereas no high-bandwidth modulation was shown in Ref.~\cite{churaevHeterogeneouslyIntegratedLithium2023}. 

We can also compare these results with hybrid LN/SiN modulators made using a different approach~\cite{boyntonHeterogeneouslyIntegratedSilicon2020}. For a modulator of length 5~mm, Ref.~\cite{boyntonHeterogeneouslyIntegratedSilicon2020} demonstrated about 31 GHz 3-dB EO bandwidth and $V_\pi L = 6.7$ V.cm. Our results show a large improvement in bandwidth from 31 GHz to 111 GHz, and a reduction (i.e., improvement) in $V_\pi L$ from 6.7 to 3.8 V.cm. 

At a different wavelength band, around 1310 nm, we have reported high-bandwidth hybrid TFLN/SiN modulators that achieved $V_\pi L = 2.7$ V.cm for $G = 5.5\ \mathrm{\mu m}$~\cite{mereImprovedFabricationScalable2023}. While a direct comparison across different wavelength bands is difficult, we can use Eq.~(\ref{eq-VpiL}) and scale the earlier results to approximately $V_\pi L = 2.9$ V.cm (for $G = 5 \ \mathrm{\mu m}$ at $1.55\ \mathrm{\mu m}$ wavelengths). The $V_\pi L$ values of the modulators reported here are about 30\% higher. However, the modulators made using the single-layer SiN design of Ref.~\cite{mereImprovedFabricationScalable2023} could not be used at 1550 nm since the insertion loss was too high to be measured (more than 30 dB). This major problem was overcome by using the bilayer SiN design in this work which ``pulls'' the optical mode below the bonded surface in the transition segment.

Finally, we emphasize that this work incorporates high-performance EOMs with an established low-loss passive photonics platform in a modular way. Designs, materials, and fabrication processes were selected to not adversely impact previously-established components, and to lower overall risk. By relying on foundry processing, we can anticipate iterative improvements of performance and scalability to larger PICs. These advances can benefit a wide range of users across many application areas, including fiber and free-space communications; analog waveform acquisition, generation, and processing; sensing; displays and beamforming; and quantum optics technologies. 

\section{Conclusion}
In conclusion, we have demonstrated hybrid LN/SiN EO modulators integrated with low-loss SiN waveguides made in a foundry process starting with standard (low-resistivity) Si handle wafers. The optical waveguide routing, splitting, and tapering was implemented using the SiN waveguides and no waveguide patterning was necessary on the LN layer. These EOMs offer an attractive combination of device performance metrics, with extinction ratio greater than 30~dB, about 3.8~dB on-chip insertion loss, a low-frequency (1~kHz--1~MHz) half-wave voltage-length product ($V_\pi L$) around 3.8~V.cm, and a high-frequency 3-dB EO modulation bandwidth exceeding 110~GHz. This combination of parameters exceeds the performance of earlier hybrid LN modulators. These results show that scalable and cost-effective wafer-scale processing can be used to add high-performance EOM capability to a low-loss SiN-based passive photonics platform.

\vspace{12 pt}
\noindent \textbf{Acknowledgments:} This research was developed in part with funding from the U.S. Government. This paper describes objective technical results and analysis. The views, opinions and/or findings expressed are those of the authors alone and should not be interpreted as representing the official views or policies of the U.S. Government. 

%-----------------------------------------------------------------------

%
%-----------------------------------------------------------------------
\cleardoublepage
\setcounter{section}{0}
\setcounter{equation}{0}
\setcounter{page}{1}
\setcounter{figure}{0}
\renewcommand{\thepage}{S\arabic{page}}
\renewcommand{\thesection}{S\arabic{section}}
\renewcommand{\theequation}{S\arabic{equation}}
\renewcommand{\thefigure}{S\arabic{figure}}

\onecolumngrid
\begin{center}
{\Large \bfseries Supplementary Information}
\end{center}
\twocolumngrid
%-----------------------------------------------------------------------
Cross-sectional images of the device structure are shown in Section~\ref{SI-section-cross-section}. A representative optical transmission spectrum of the asymmetric Mach-Zehnder interferometer is shown in Section~\ref{SI-section-spectrum}. Section~\ref{SI-section-EORgraphs} shows a composite figure of the electro-optic response (EOR) for the three design types across three different chips. A discussion of the fitting function for the full-band electro-optic response (EOR) is presented in Section~\ref{SI-section-EOR}.

%----------------------------------------------
\section{Cross-sectional Images}
\label{SI-section-cross-section}

\begin{figure*}[tbh]
\centering
\includegraphics[width=0.7\linewidth, clip=true]{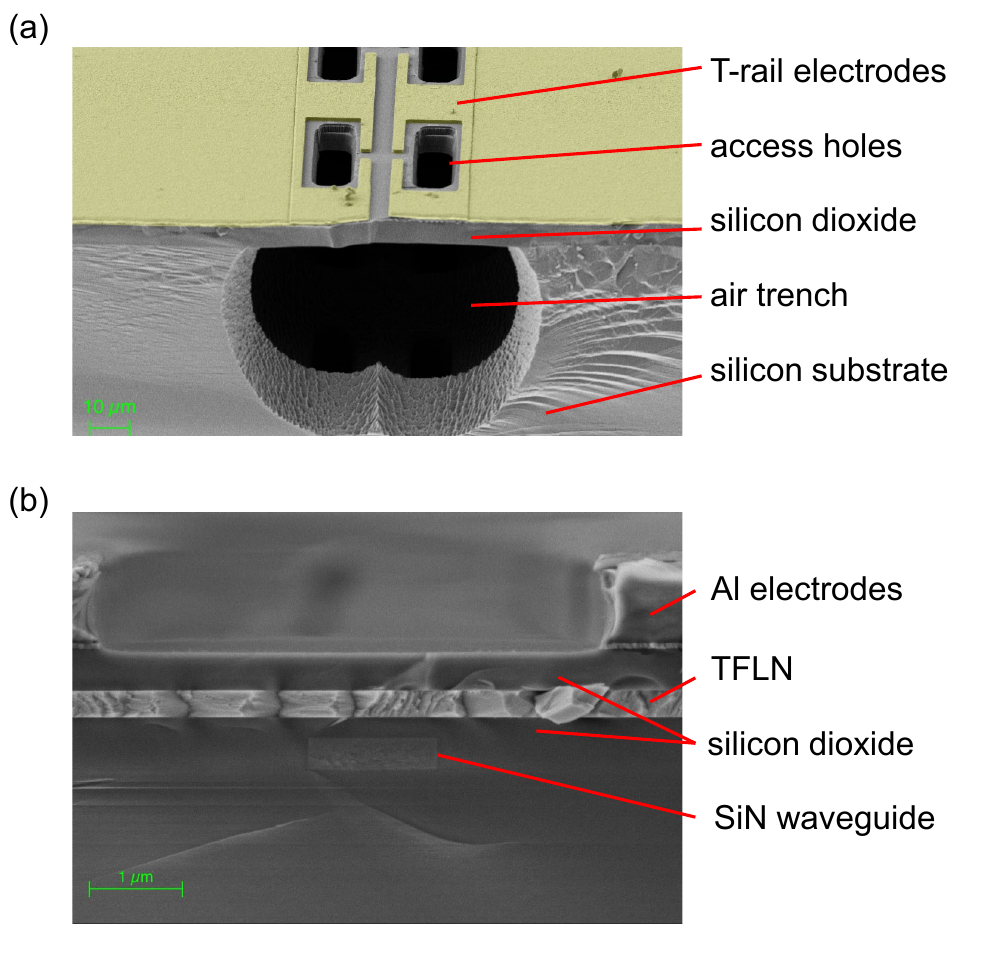}
\caption{(a) Cross-sectional scanning-electron microscope (SEM) image showing the trenches etched into the Si substrate to lower the propagation loss of the traveling-wave microwave mode. (b) Higher-resolution SEM image of a test chip showing the SiN2 waveguide layer and the TFLN bonded region.} 
\label{fig-S-cross-section}
\end{figure*}

Figure~\ref{fig-S-cross-section} shows tilted-angle scanning-electron microscope (SEM) images of the cross-section of the EOM device with an air trench. In panel (a), the false-colored SEM image shows the trench access holes and the trench etched into the silicon (Si) substrate. The scalebar at the bottom left corner of this image shows a length of $10\ \mathrm{\mu m}$. As discussed in the main text, removing some of the low-resistivity Si substrate (resistivity about $10 \ \mathrm{\Omega .cm}$) lowers the RF propagation loss of the traveling-wave electrodes and supports high-frequency electro-optic (EO) modulation. 

In panel (b), the image shows the layers that define the optical hybrid mode, consisting of a SiN strip waveguide, a thin layer of separation oxide, and the thin-film lithium niobate (TFLN) slab with oxide above it. A portion of the metal electrode is visible at the right-hand side edge of the image.  The scalebar at the bottom left corner of this image shows a length of $1\ \mathrm{\mu m}$. The rough edge was seen because of unpolished manual cleaving of the chip. For the device chips, the waveguide-fiber coupler at the edge of the chip was defined using a SiN1 layer component which has been previously studied by LIGENTEC and offers about 1 dB per facet coupling loss to lensed fibers with a mode-field diameter of about $2.5\ \mathrm{\mu m}$ at 1550 nm (LIGENTEC unpublished data). 

%-----------------------------------------------------------
%-----------------------------------------------------------

%-----------------------------------------------------------
\begin{figure*}[tbh]
\centering
\includegraphics[width=0.8\linewidth, clip=true]{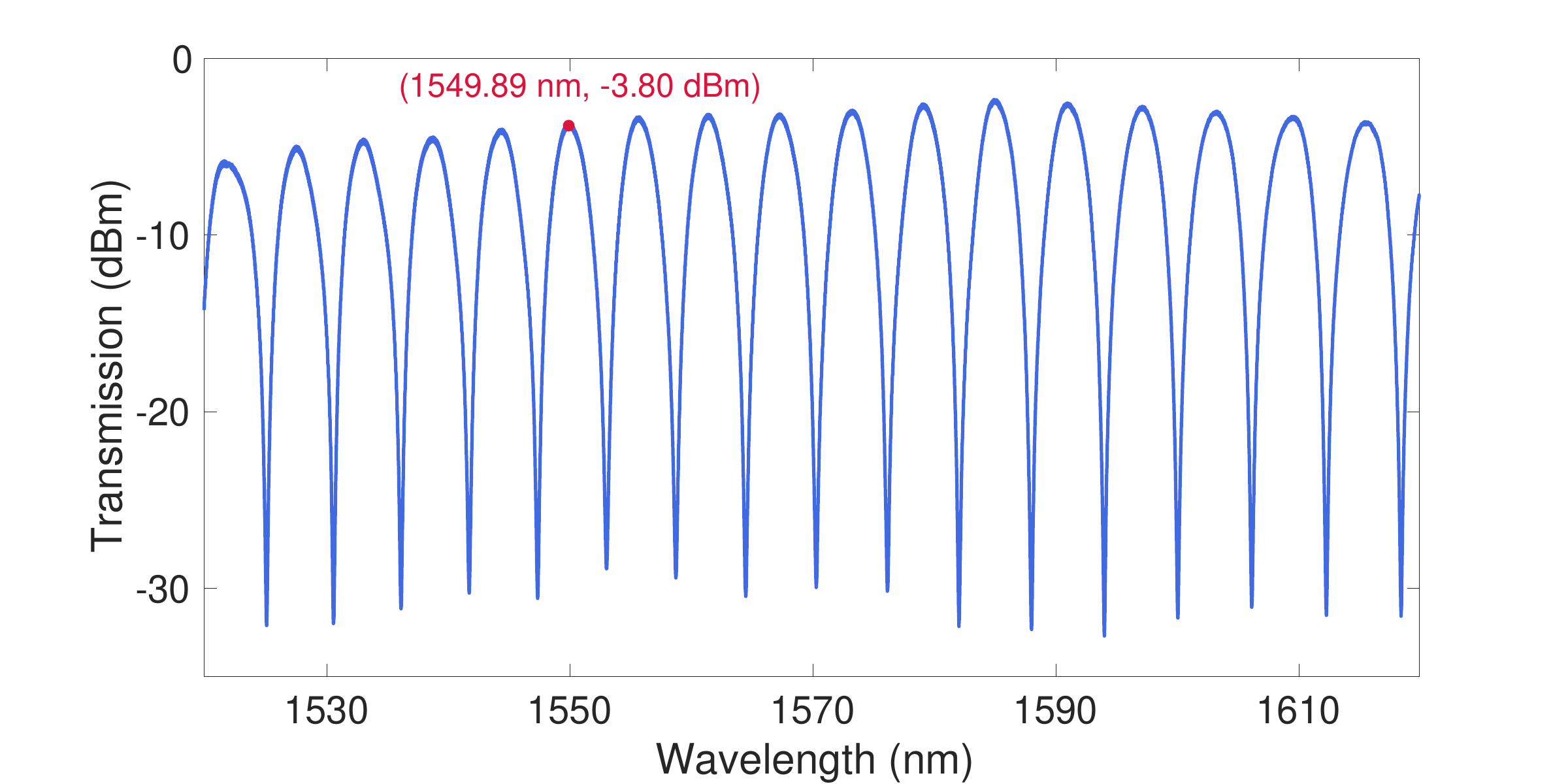}
\caption{The optical transmission of the asymmetric Mach-Zehnder interferometer structure that forms part of the EOM devices.} 
\label{fig-S-opticaltransmission}
\end{figure*}
%-----------------------------------------------------------
%----------------------------------------------
\section{Optical Transmission Spectrum}
\label{SI-section-spectrum}
Figure~\ref{fig-S-opticaltransmission} shows the optical transmission spectrum between 1520 nm and 1620 nm of a representative EOM device without applying any voltage to the electrodes. Similar data was obtained for all the test devices in this batch (12 devices in total). 

\textit{Biasing:}~The path length difference (PLD) segment in the asymmetric Mach-Zehnder interferometer (MZI) causes the periodicity seen in the transmission spectrum. We used this information to determine the optimal wavelength at which to test the performance of the EOM, by selecting a wavelength at which the transmission is biased 3~dB below the peak transmission. This corresponds to the linear-most portion of the inherently-nonlinear modulation transfer function (MTF) of a MZI-based modulator device~\cite{ghioneSemiconductorDevicesHighspeed2009}. 

\textit{Optical Insertion Loss:}~The raw data was normalized so that 0 dB corresponds to no additional insertion loss compared to a straight-through waveguide only in the SiN1 layer also fabricated on the same chip. Both devices used the same type of waveguide-fiber edge couplers and have the same physical length on the chip. We first measured the EOM device, found the wavelength for a transmission peak near 1550 nm, and then measured the transmission of the straight-through waveguide at the same wavelength. Subtracting the first number from the second, we conclude that the EOM transmission was 3.80 dB lower than the waveguide transmission at 1549.89 nm, which therefore gives the on-chip insertion loss of this EOM device. As mentioned in the main text, the measured average on-chip IL (and one standard deviation) for test devices across four chips in the three different MZM designs were: 2.95 dB (1.43 dB) for Design 1, 6.43 dB (2.75 dB) for Design 2, and 3.77 dB (0.83 dB). 

\textit{Extinction Ratio:}~An average extinction ratio (ER) of about 30 dB was measured across a wide range of wavelengths around 1550 nm. The high ER shows that the input and output 3-dB couplers were well-balanced, and the losses in both arms of the push-pull MZI structure were also well-balanced after fabrication was completed. The PLD segment was defined in the SiN1 layer, as were the 3-dB multi-mode interference (MMI) couplers at the input and output segments of the MZI. The design and performance of these components will not change if there are minor variations in the SiN2, oxide or TFLN thicknesses across the wafer. The SiN1 process is already quite mature at LIGENTEC.    

%----------------------------------------------
\section{Electro-Optic Response (EOR) Measurements}
\label{SI-section-EORgraphs}
\begin{figure*}[tbh]
\centering
\includegraphics[width=\linewidth, clip=true]{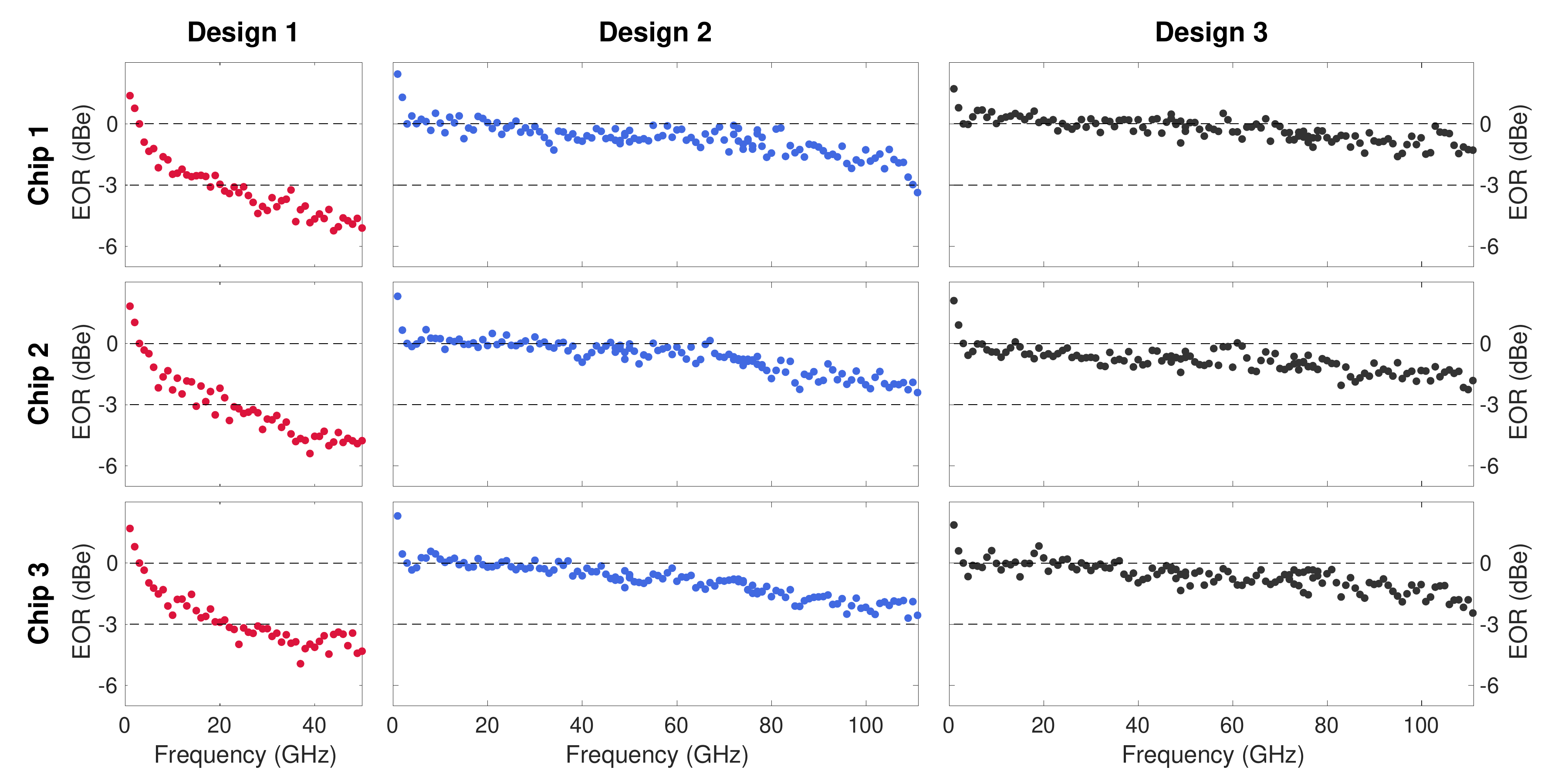}
\caption{The electro-optic response (EOR) for Design 1 (red), Design 2 (blue), and Design 3 (black) on three different chips. Each data set is individually normalized to the EOR value at 3 GHz.} 
\label{fig-S-allEOR}
\end{figure*}

Figure~\ref{fig-S-allEOR} shows the EOR data measured for EOMs with Designs 1, 2 and 3 on three test chips. These chips were singulated from the wafer by dicing and have slightly different layer thicknesses and feature dimensions. However, the EOR performance of each design was similar. 

%-----------------------------------------------------------
%-----------------------------------------------------------

The main paper discusses the normalization of the EOR to its value at 1 GHz, or to the average value at the lower frequencies between 1 GHz and 3 GHz, or alternative approaches. Here, for simplicity, each set of EOR data was normalized to its value at 3 GHz. 

Design 1 has the fastest roll-off and drops off by more than 3 dB well before 50 GHz. (Thus, there was no need for the RF multipliers to measure the 3-dB modulation bandwidth of this design, which is around 20 GHz, and the horizontal frequency axis of the first column in Fig.~\ref{fig-S-allEOR} is limited to 50 GHz.) Designs 2 and 3 have a 3-dB modulation bandwidth around 110 GHz or higher. Up to the measurement limit of our apparatus (111 GHz), we do not see evidence of a sharp roll-off in the EOR below the horizontal dashed lines indicating the 3-dB range, which is already visible in Design 1 below 50 GHz. 

The EOR shown in the top right-hand side corner (Chip 1, Design 3) is shown and discussed in the main paper [Fig. 4(c) and 4(d)].  

%----------------------------------------------
\section{Electro-Optic Response (EOR) Theory}
\label{SI-section-EOR}
As stated in the main manuscript, we used a modulation-sideband method for measuring the electro-optic response (EOR). A comparison against an EOR measurement made using a lightwave component analyzer (LCA) was discussed in our earlier work~\cite{valdezIntegratedCbandSiliconlithium2023}.

The black curve in Fig.~4(c) [main paper] is the fitted EO modulation response which was calculated using a traveling-wave model of the EO interaction~\cite{ghioneSemiconductorDevicesHighspeed2009}: 

\begin{widetext}
\begin{equation} \label{eq_EOR}
m(\omega)=\frac{R_{L}+R_{G}}{R_{L}}\left|\frac{Z_\mathrm{in}}{Z_\mathrm{in}+Z_\mathrm{G}}\right|\left|\frac{(Z_\mathrm{L}+Z_\mathrm{c})F(u_{+})+(Z_\mathrm{L}-Z_\mathrm{c})F(u_{-})}{(Z_\mathrm{L}+Z_\mathrm{c})\exp[\gamma_\mathrm{m}L]+(Z_\mathrm{L}-Z_\mathrm{c})\exp[-\gamma_\mathrm{m}L]}\right|,
\end{equation}
\end{widetext}
with the following definitions:
\begin{subequations}
\begin{equation}
Z_\mathrm{in}=Z_\mathrm{c}\frac{Z_\mathrm{L}+Z_\mathrm{c}\mathrm{tanh}(\gamma_\mathrm{m}L)}{Z_\mathrm{c}+Z_\mathrm{L}\mathrm{tanh}(\gamma_\mathrm{m}L)},
\end{equation}
\begin{equation} \label{eq_gm}
\gamma_\mathrm{m}=\alpha_\mathrm{m}+\frac{j\omega}{c}n_\mathrm{m}.
\end{equation}
\begin{equation} \label{eq_Fu}
F(u_{\pm}(P))=\frac{1-\exp[u_{\pm}]}{u_{\pm}}
\end{equation}
\begin{equation} \label{eq_upm}
u_{\pm}(P)=\pm\alpha_\mathrm{m}L+\frac{j\omega}{c}(\pm n_\mathrm{m}-n_\mathrm{g})L.
\end{equation}
\end{subequations}
where $R_{\{{L,G}\}}$ are the load and generator resistances, $Z_{\{{\textrm{in},L,G}\}}$ are the input, load, and generator impedances, $\alpha_\mathrm{m}$ is the RF propagation loss, $n_\mathrm{m}$ is the RF effective index, and $\gamma_\textrm{m}$ is the complex propagation constant of the RF wave along the transmission line and $L_\mathrm{ps}$ is the phase-shifter interaction length.

Using this model, we use an iterative nonlinear programming solver based on the Nelder-Mead Simplex method (implemented in MATLAB as ``fminsearch'') to fit the measured data, with the following assumptions: $R_{L,G} = 50\ \Omega$, $L_\mathrm{ps}=0.6$ cm, optical group index $n_g = 2.0854$ (obtained from a measurement). We assume that the microwave loss increases with the square-root of the modulation RF frequency $f$, $\alpha_m(f) =\alpha_{m0} f^{1/2}$, where $\alpha_{m0}$ is the skin-effect loss coefficient and is a fitting parameter to be found by optimization. This scaling behavior is a good, but not perfect, fit to the data, as we have reported elsewhere on a different batch of microchips~\cite{valdezIntegratedCbandSiliconlithium2023}. However, the exact value of the scaling exponent (between $1/4$ and $1/2$ in some frequency bands) does not greatly affect either the fitted line or the conclusions. Since Eq.~(\ref{eq_EOR}) is nonlinear, the curve may not be the unique fit to the data, but we have checked that the same solution was obtained with slightly different starting assumptions for the fitting parameters. 

The black curve in Fig.~4(d) [main paper] is the mathematical transformation using Eq.~(1) [main paper] of the black line in Fig.~4(c) [main paper]. 


\begin{thebibliography}{45}%
\makeatletter
\providecommand \@ifxundefined [1]{%
 \@ifx{#1\undefined}
}%
\providecommand \@ifnum [1]{%
 \ifnum #1\expandafter \@firstoftwo
 \else \expandafter \@secondoftwo
 \fi
}%
\providecommand \@ifx [1]{%
 \ifx #1\expandafter \@firstoftwo
 \else \expandafter \@secondoftwo
 \fi
}%
\providecommand \natexlab [1]{#1}%
\providecommand \enquote  [1]{``#1''}%
\providecommand \bibnamefont  [1]{#1}%
\providecommand \bibfnamefont [1]{#1}%
\providecommand \citenamefont [1]{#1}%
\providecommand \href@noop [0]{\@secondoftwo}%
\providecommand \href [0]{\begingroup \@sanitize@url \@href}%
\providecommand \@href[1]{\@@startlink{#1}\@@href}%
\providecommand \@@href[1]{\endgroup#1\@@endlink}%
\providecommand \@sanitize@url [0]{\catcode `\\12\catcode `\$12\catcode
  `\&12\catcode `\#12\catcode `\^12\catcode `\_12\catcode `\%12\relax}%
\providecommand \@@startlink[1]{}%
\providecommand \@@endlink[0]{}%
\providecommand \url  [0]{\begingroup\@sanitize@url \@url }%
\providecommand \@url [1]{\endgroup\@href {#1}{\urlprefix }}%
\providecommand \urlprefix  [0]{URL }%
\providecommand \Eprint [0]{\href }%
\providecommand \doibase [0]{https://doi.org/}%
\providecommand \selectlanguage [0]{\@gobble}%
\providecommand \bibinfo  [0]{\@secondoftwo}%
\providecommand \bibfield  [0]{\@secondoftwo}%
\providecommand \translation [1]{[#1]}%
\providecommand \BibitemOpen [0]{}%
\providecommand \bibitemStop [0]{}%
\providecommand \bibitemNoStop [0]{.\EOS\space}%
\providecommand \EOS [0]{\spacefactor3000\relax}%
\providecommand \BibitemShut  [1]{\csname bibitem#1\endcsname}%
\let\auto@bib@innerbib\@empty
%</preamble>
\bibitem [{\citenamefont {Wooten}\ \emph {et~al.}(2000)\citenamefont {Wooten},
  \citenamefont {Kissa}, \citenamefont {{Yi-Yan}}, \citenamefont {Murphy},
  \citenamefont {Lafaw}, \citenamefont {Hallemeier}, \citenamefont {Maack},
  \citenamefont {Attanasio}, \citenamefont {Fritz}, \citenamefont {McBrien},\
  and\ \citenamefont {Bossi}}]{wootenReviewLithiumNiobate2000}%
  \BibitemOpen
  \bibfield  {author} {\bibinfo {author} {\bibfnamefont {E.~L.}\ \bibnamefont
  {Wooten}}, \bibinfo {author} {\bibfnamefont {K.~M.}\ \bibnamefont {Kissa}},
  \bibinfo {author} {\bibfnamefont {A.}~\bibnamefont {{Yi-Yan}}}, \bibinfo
  {author} {\bibfnamefont {E.~J.}\ \bibnamefont {Murphy}}, \bibinfo {author}
  {\bibfnamefont {D.~A.}\ \bibnamefont {Lafaw}}, \bibinfo {author}
  {\bibfnamefont {P.~F.}\ \bibnamefont {Hallemeier}}, \bibinfo {author}
  {\bibfnamefont {D.}~\bibnamefont {Maack}}, \bibinfo {author} {\bibfnamefont
  {D.~V.}\ \bibnamefont {Attanasio}}, \bibinfo {author} {\bibfnamefont {D.~J.}\
  \bibnamefont {Fritz}}, \bibinfo {author} {\bibfnamefont {G.~J.}\ \bibnamefont
  {McBrien}},\ and\ \bibinfo {author} {\bibfnamefont {D.~E.}\ \bibnamefont
  {Bossi}},\ }\bibfield  {title} {\bibinfo {title} {A review of lithium niobate
  modulators for fiber-optic communications systems},\ }\href
  {https://doi.org/10.1109/2944.826874} {\bibfield  {journal} {\bibinfo
  {journal} {IEEE Journal of Selected Topics in Quantum Electronics}\ }\textbf
  {\bibinfo {volume} {6}},\ \bibinfo {pages} {69} (\bibinfo {year}
  {2000})}\BibitemShut {NoStop}%
\bibitem [{\citenamefont {Howerton}\ and\ \citenamefont
  {Burns}(2002)}]{howertonBroadbandTravelingWave2002}%
  \BibitemOpen
  \bibfield  {author} {\bibinfo {author} {\bibfnamefont {M.~M.}\ \bibnamefont
  {Howerton}}\ and\ \bibinfo {author} {\bibfnamefont {W.~K.}\ \bibnamefont
  {Burns}},\ }\bibfield  {title} {\bibinfo {title} {Broadband traveling wave
  modulators in {{LiNbO3}}},\ }in\ \href@noop {} {\emph {\bibinfo {booktitle}
  {{{RF Photonic Technology}} in {{Optical Fiber Links}}}}},\ \bibinfo {editor}
  {edited by\ \bibinfo {editor} {\bibfnamefont {W.~S.~C.}\ \bibnamefont
  {Chang}}}\ (\bibinfo  {publisher} {Cambridge University},\ \bibinfo {year}
  {2002})\BibitemShut {NoStop}%
\bibitem [{\citenamefont {Yu}\ \emph {et~al.}(2014)\citenamefont {Yu},
  \citenamefont {Ying}, \citenamefont {Pantouvaki}, \citenamefont
  {Van~Campenhout}, \citenamefont {Absil}, \citenamefont {Hao}, \citenamefont
  {Yang},\ and\ \citenamefont {Jiang}}]{yuTradeoffOpticalModulation2014}%
  \BibitemOpen
  \bibfield  {author} {\bibinfo {author} {\bibfnamefont {H.}~\bibnamefont
  {Yu}}, \bibinfo {author} {\bibfnamefont {D.}~\bibnamefont {Ying}}, \bibinfo
  {author} {\bibfnamefont {M.}~\bibnamefont {Pantouvaki}}, \bibinfo {author}
  {\bibfnamefont {J.}~\bibnamefont {Van~Campenhout}}, \bibinfo {author}
  {\bibfnamefont {P.}~\bibnamefont {Absil}}, \bibinfo {author} {\bibfnamefont
  {Y.}~\bibnamefont {Hao}}, \bibinfo {author} {\bibfnamefont {J.}~\bibnamefont
  {Yang}},\ and\ \bibinfo {author} {\bibfnamefont {X.}~\bibnamefont {Jiang}},\
  }\bibfield  {title} {\bibinfo {title} {Trade-off between optical modulation
  amplitude and modulation bandwidth of silicon micro-ring modulators},\ }\href
  {https://doi.org/10.1364/OE.22.015178} {\bibfield  {journal} {\bibinfo
  {journal} {Optics Express}\ }\textbf {\bibinfo {volume} {22}},\ \bibinfo
  {pages} {15178} (\bibinfo {year} {2014})}\BibitemShut {NoStop}%
\bibitem [{\citenamefont
  {Witzens}(2018)}]{witzensHighSpeedSiliconPhotonics2018}%
  \BibitemOpen
  \bibfield  {author} {\bibinfo {author} {\bibfnamefont {J.}~\bibnamefont
  {Witzens}},\ }\bibfield  {title} {\bibinfo {title} {High-{{Speed Silicon
  Photonics Modulators}}},\ }\href {https://doi.org/10.1109/JPROC.2018.2877636}
  {\bibfield  {journal} {\bibinfo  {journal} {Proceedings of the IEEE}\
  }\textbf {\bibinfo {volume} {106}},\ \bibinfo {pages} {2158} (\bibinfo {year}
  {2018})}\BibitemShut {NoStop}%
\bibitem [{\citenamefont {Jain}\ \emph {et~al.}(2019)\citenamefont {Jain},
  \citenamefont {Hosseinzadeh}, \citenamefont {Wu}, \citenamefont {Tsang},
  \citenamefont {Helkey}, \citenamefont {Bowers},\ and\ \citenamefont
  {Buckwalter}}]{jainHighSpurFreeDynamic2019}%
  \BibitemOpen
  \bibfield  {author} {\bibinfo {author} {\bibfnamefont {A.}~\bibnamefont
  {Jain}}, \bibinfo {author} {\bibfnamefont {N.}~\bibnamefont {Hosseinzadeh}},
  \bibinfo {author} {\bibfnamefont {X.}~\bibnamefont {Wu}}, \bibinfo {author}
  {\bibfnamefont {H.~K.}\ \bibnamefont {Tsang}}, \bibinfo {author}
  {\bibfnamefont {R.}~\bibnamefont {Helkey}}, \bibinfo {author} {\bibfnamefont
  {J.~E.}\ \bibnamefont {Bowers}},\ and\ \bibinfo {author} {\bibfnamefont
  {J.~F.}\ \bibnamefont {Buckwalter}},\ }\bibfield  {title} {\bibinfo {title}
  {A {{High Spur-Free Dynamic Range Silicon DC Kerr Ring Modulator}} for {{RF
  Applications}}},\ }\href {https://doi.org/10.1109/JLT.2019.2913638}
  {\bibfield  {journal} {\bibinfo  {journal} {Journal of Lightwave Technology}\
  }\textbf {\bibinfo {volume} {37}},\ \bibinfo {pages} {3261} (\bibinfo {year}
  {2019})}\BibitemShut {NoStop}%
\bibitem [{\citenamefont {Shekhar}\ \emph {et~al.}(2024)\citenamefont
  {Shekhar}, \citenamefont {Bogaerts}, \citenamefont {Chrostowski},
  \citenamefont {Bowers}, \citenamefont {Hochberg}, \citenamefont {Soref},\
  and\ \citenamefont {Shastri}}]{shekharRoadmappingNextGeneration2024}%
  \BibitemOpen
  \bibfield  {author} {\bibinfo {author} {\bibfnamefont {S.}~\bibnamefont
  {Shekhar}}, \bibinfo {author} {\bibfnamefont {W.}~\bibnamefont {Bogaerts}},
  \bibinfo {author} {\bibfnamefont {L.}~\bibnamefont {Chrostowski}}, \bibinfo
  {author} {\bibfnamefont {J.~E.}\ \bibnamefont {Bowers}}, \bibinfo {author}
  {\bibfnamefont {M.}~\bibnamefont {Hochberg}}, \bibinfo {author}
  {\bibfnamefont {R.}~\bibnamefont {Soref}},\ and\ \bibinfo {author}
  {\bibfnamefont {B.~J.}\ \bibnamefont {Shastri}},\ }\bibfield  {title}
  {\bibinfo {title} {Roadmapping the next generation of silicon photonics},\
  }\href {https://doi.org/10.1038/s41467-024-44750-0} {\bibfield  {journal}
  {\bibinfo  {journal} {Nature Communications}\ }\textbf {\bibinfo {volume}
  {15}},\ \bibinfo {pages} {751} (\bibinfo {year} {2024})}\BibitemShut
  {NoStop}%
\bibitem [{\citenamefont {Poberaj}\ \emph {et~al.}(2012)\citenamefont
  {Poberaj}, \citenamefont {Hu}, \citenamefont {Sohler},\ and\ \citenamefont
  {G{\"u}nter}}]{poberajLithiumNiobateInsulator2012}%
  \BibitemOpen
  \bibfield  {author} {\bibinfo {author} {\bibfnamefont {G.}~\bibnamefont
  {Poberaj}}, \bibinfo {author} {\bibfnamefont {H.}~\bibnamefont {Hu}},
  \bibinfo {author} {\bibfnamefont {W.}~\bibnamefont {Sohler}},\ and\ \bibinfo
  {author} {\bibfnamefont {P.}~\bibnamefont {G{\"u}nter}},\ }\bibfield  {title}
  {\bibinfo {title} {Lithium niobate on insulator ({{LNOI}}) for micro-photonic
  devices},\ }\href {https://doi.org/10.1002/lpor.201100035} {\bibfield
  {journal} {\bibinfo  {journal} {Laser \& Photonics Reviews}\ }\textbf
  {\bibinfo {volume} {6}},\ \bibinfo {pages} {488} (\bibinfo {year}
  {2012})}\BibitemShut {NoStop}%
\bibitem [{\citenamefont {Wang}\ \emph {et~al.}(2018)\citenamefont {Wang},
  \citenamefont {Zhang}, \citenamefont {Chen}, \citenamefont {Bertrand},
  \citenamefont {{Shams-Ansari}}, \citenamefont {Chandrasekhar}, \citenamefont
  {Winzer},\ and\ \citenamefont {Lon{\v
  c}ar}}]{wangIntegratedLithiumNiobate2018}%
  \BibitemOpen
  \bibfield  {author} {\bibinfo {author} {\bibfnamefont {C.}~\bibnamefont
  {Wang}}, \bibinfo {author} {\bibfnamefont {M.}~\bibnamefont {Zhang}},
  \bibinfo {author} {\bibfnamefont {X.}~\bibnamefont {Chen}}, \bibinfo {author}
  {\bibfnamefont {M.}~\bibnamefont {Bertrand}}, \bibinfo {author}
  {\bibfnamefont {A.}~\bibnamefont {{Shams-Ansari}}}, \bibinfo {author}
  {\bibfnamefont {S.}~\bibnamefont {Chandrasekhar}}, \bibinfo {author}
  {\bibfnamefont {P.}~\bibnamefont {Winzer}},\ and\ \bibinfo {author}
  {\bibfnamefont {M.}~\bibnamefont {Lon{\v c}ar}},\ }\bibfield  {title}
  {\bibinfo {title} {Integrated lithium niobate electro-optic modulators
  operating at {{CMOS-compatible}} voltages},\ }\href
  {https://doi.org/10.1038/s41586-018-0551-y} {\bibfield  {journal} {\bibinfo
  {journal} {Nature}\ }\textbf {\bibinfo {volume} {562}},\ \bibinfo {pages}
  {101} (\bibinfo {year} {2018})}\BibitemShut {NoStop}%
\bibitem [{\citenamefont {Weigel}\ \emph {et~al.}(2018)\citenamefont {Weigel},
  \citenamefont {Zhao}, \citenamefont {Fang}, \citenamefont {{Al-Rubaye}},
  \citenamefont {Trotter}, \citenamefont {Hood}, \citenamefont {Mudrick},
  \citenamefont {Dallo}, \citenamefont {Pomerene}, \citenamefont {Starbuck},
  \citenamefont {DeRose}, \citenamefont {Lentine}, \citenamefont {Rebeiz},\
  and\ \citenamefont {Mookherjea}}]{weigelBondedThinFilm2018a}%
  \BibitemOpen
  \bibfield  {author} {\bibinfo {author} {\bibfnamefont {P.~O.}\ \bibnamefont
  {Weigel}}, \bibinfo {author} {\bibfnamefont {J.}~\bibnamefont {Zhao}},
  \bibinfo {author} {\bibfnamefont {K.}~\bibnamefont {Fang}}, \bibinfo {author}
  {\bibfnamefont {H.}~\bibnamefont {{Al-Rubaye}}}, \bibinfo {author}
  {\bibfnamefont {D.}~\bibnamefont {Trotter}}, \bibinfo {author} {\bibfnamefont
  {D.}~\bibnamefont {Hood}}, \bibinfo {author} {\bibfnamefont {J.}~\bibnamefont
  {Mudrick}}, \bibinfo {author} {\bibfnamefont {C.}~\bibnamefont {Dallo}},
  \bibinfo {author} {\bibfnamefont {A.~T.}\ \bibnamefont {Pomerene}}, \bibinfo
  {author} {\bibfnamefont {A.~L.}\ \bibnamefont {Starbuck}}, \bibinfo {author}
  {\bibfnamefont {C.~T.}\ \bibnamefont {DeRose}}, \bibinfo {author}
  {\bibfnamefont {A.~L.}\ \bibnamefont {Lentine}}, \bibinfo {author}
  {\bibfnamefont {G.}~\bibnamefont {Rebeiz}},\ and\ \bibinfo {author}
  {\bibfnamefont {S.}~\bibnamefont {Mookherjea}},\ }\bibfield  {title}
  {\bibinfo {title} {Bonded thin film lithium niobate modulator on a silicon
  photonics platform exceeding 100 {{GHz}} 3-{{dB}} electrical modulation
  bandwidth},\ }\href {https://doi.org/10.1364/OE.26.023728} {\bibfield
  {journal} {\bibinfo  {journal} {Optics Express}\ }\textbf {\bibinfo {volume}
  {26}},\ \bibinfo {pages} {23728} (\bibinfo {year} {2018})}\BibitemShut
  {NoStop}%
\bibitem [{\citenamefont {Mercante}\ \emph {et~al.}(2018)\citenamefont
  {Mercante}, \citenamefont {Shi}, \citenamefont {Yao}, \citenamefont {Xie},
  \citenamefont {Weikle},\ and\ \citenamefont
  {Prather}}]{mercanteThinFilmLithium2018}%
  \BibitemOpen
  \bibfield  {author} {\bibinfo {author} {\bibfnamefont {A.~J.}\ \bibnamefont
  {Mercante}}, \bibinfo {author} {\bibfnamefont {S.}~\bibnamefont {Shi}},
  \bibinfo {author} {\bibfnamefont {P.}~\bibnamefont {Yao}}, \bibinfo {author}
  {\bibfnamefont {L.}~\bibnamefont {Xie}}, \bibinfo {author} {\bibfnamefont
  {R.~M.}\ \bibnamefont {Weikle}},\ and\ \bibinfo {author} {\bibfnamefont
  {D.~W.}\ \bibnamefont {Prather}},\ }\bibfield  {title} {\bibinfo {title}
  {Thin film lithium niobate electro-optic modulator with terahertz operating
  bandwidth},\ }\href {https://doi.org/10.1364/OE.26.014810} {\bibfield
  {journal} {\bibinfo  {journal} {Optics Express}\ }\textbf {\bibinfo {volume}
  {26}},\ \bibinfo {pages} {14810} (\bibinfo {year} {2018})}\BibitemShut
  {NoStop}%
\bibitem [{\citenamefont {He}\ \emph {et~al.}(2019)\citenamefont {He},
  \citenamefont {Xu}, \citenamefont {Ren}, \citenamefont {Jian}, \citenamefont
  {Ruan}, \citenamefont {Xu}, \citenamefont {Gao}, \citenamefont {Sun},
  \citenamefont {Wen}, \citenamefont {Zhou}, \citenamefont {Liu}, \citenamefont
  {Guo}, \citenamefont {Chen}, \citenamefont {Yu}, \citenamefont {Liu},\ and\
  \citenamefont {Cai}}]{heHighperformanceHybridSilicon2019}%
  \BibitemOpen
  \bibfield  {author} {\bibinfo {author} {\bibfnamefont {M.}~\bibnamefont
  {He}}, \bibinfo {author} {\bibfnamefont {M.}~\bibnamefont {Xu}}, \bibinfo
  {author} {\bibfnamefont {Y.}~\bibnamefont {Ren}}, \bibinfo {author}
  {\bibfnamefont {J.}~\bibnamefont {Jian}}, \bibinfo {author} {\bibfnamefont
  {Z.}~\bibnamefont {Ruan}}, \bibinfo {author} {\bibfnamefont {Y.}~\bibnamefont
  {Xu}}, \bibinfo {author} {\bibfnamefont {S.}~\bibnamefont {Gao}}, \bibinfo
  {author} {\bibfnamefont {S.}~\bibnamefont {Sun}}, \bibinfo {author}
  {\bibfnamefont {X.}~\bibnamefont {Wen}}, \bibinfo {author} {\bibfnamefont
  {L.}~\bibnamefont {Zhou}}, \bibinfo {author} {\bibfnamefont {L.}~\bibnamefont
  {Liu}}, \bibinfo {author} {\bibfnamefont {C.}~\bibnamefont {Guo}}, \bibinfo
  {author} {\bibfnamefont {H.}~\bibnamefont {Chen}}, \bibinfo {author}
  {\bibfnamefont {S.}~\bibnamefont {Yu}}, \bibinfo {author} {\bibfnamefont
  {L.}~\bibnamefont {Liu}},\ and\ \bibinfo {author} {\bibfnamefont
  {X.}~\bibnamefont {Cai}},\ }\bibfield  {title} {\bibinfo {title}
  {High-performance hybrid silicon and lithium niobate {{Mach}}--{{Zehnder}}
  modulators for 100 {{Gbits-1}} and beyond},\ }\href
  {https://doi.org/10.1038/s41566-019-0378-6} {\bibfield  {journal} {\bibinfo
  {journal} {Nature Photonics}\ }\textbf {\bibinfo {volume} {13}},\ \bibinfo
  {pages} {359} (\bibinfo {year} {2019})}\BibitemShut {NoStop}%
\bibitem [{\citenamefont {Chen}\ \emph {et~al.}(2014)\citenamefont {Chen},
  \citenamefont {Xu}, \citenamefont {Wood},\ and\ \citenamefont
  {Reano}}]{chenHybridSiliconLithium2014}%
  \BibitemOpen
  \bibfield  {author} {\bibinfo {author} {\bibfnamefont {L.}~\bibnamefont
  {Chen}}, \bibinfo {author} {\bibfnamefont {Q.}~\bibnamefont {Xu}}, \bibinfo
  {author} {\bibfnamefont {M.~G.}\ \bibnamefont {Wood}},\ and\ \bibinfo
  {author} {\bibfnamefont {R.~M.}\ \bibnamefont {Reano}},\ }\bibfield  {title}
  {\bibinfo {title} {Hybrid silicon and lithium niobate electro-optical ring
  modulator},\ }\href {https://doi.org/10.1364/OPTICA.1.000112} {\bibfield
  {journal} {\bibinfo  {journal} {Optica}\ }\textbf {\bibinfo {volume} {1}},\
  \bibinfo {pages} {112} (\bibinfo {year} {2014})}\BibitemShut {NoStop}%
\bibitem [{\citenamefont {Weigel}\ \emph {et~al.}(2016)\citenamefont {Weigel},
  \citenamefont {Savanier}, \citenamefont {DeRose}, \citenamefont {Pomerene},
  \citenamefont {Starbuck}, \citenamefont {Lentine}, \citenamefont {Stenger},\
  and\ \citenamefont {Mookherjea}}]{weigelLightwaveCircuitsLithium2016a}%
  \BibitemOpen
  \bibfield  {author} {\bibinfo {author} {\bibfnamefont {P.~O.}\ \bibnamefont
  {Weigel}}, \bibinfo {author} {\bibfnamefont {M.}~\bibnamefont {Savanier}},
  \bibinfo {author} {\bibfnamefont {C.~T.}\ \bibnamefont {DeRose}}, \bibinfo
  {author} {\bibfnamefont {A.~T.}\ \bibnamefont {Pomerene}}, \bibinfo {author}
  {\bibfnamefont {A.~L.}\ \bibnamefont {Starbuck}}, \bibinfo {author}
  {\bibfnamefont {A.~L.}\ \bibnamefont {Lentine}}, \bibinfo {author}
  {\bibfnamefont {V.}~\bibnamefont {Stenger}},\ and\ \bibinfo {author}
  {\bibfnamefont {S.}~\bibnamefont {Mookherjea}},\ }\bibfield  {title}
  {\bibinfo {title} {Lightwave {{Circuits}} in {{Lithium Niobate}} through
  {{Hybrid Waveguides}} with {{Silicon Photonics}}},\ }\href
  {https://doi.org/10.1038/srep22301} {\bibfield  {journal} {\bibinfo
  {journal} {Scientific Reports}\ }\textbf {\bibinfo {volume} {6}},\ \bibinfo
  {pages} {22301} (\bibinfo {year} {2016})}\BibitemShut {NoStop}%
\bibitem [{\citenamefont {Boynton}\ \emph {et~al.}(2020)\citenamefont
  {Boynton}, \citenamefont {Cai}, \citenamefont {Gehl}, \citenamefont
  {Arterburn}, \citenamefont {Dallo}, \citenamefont {Pomerene}, \citenamefont
  {Starbuck}, \citenamefont {Hood}, \citenamefont {Trotter}, \citenamefont
  {Friedmann}, \citenamefont {DeRose},\ and\ \citenamefont
  {Lentine}}]{boyntonHeterogeneouslyIntegratedSilicon2020}%
  \BibitemOpen
  \bibfield  {author} {\bibinfo {author} {\bibfnamefont {N.}~\bibnamefont
  {Boynton}}, \bibinfo {author} {\bibfnamefont {H.}~\bibnamefont {Cai}},
  \bibinfo {author} {\bibfnamefont {M.}~\bibnamefont {Gehl}}, \bibinfo {author}
  {\bibfnamefont {S.}~\bibnamefont {Arterburn}}, \bibinfo {author}
  {\bibfnamefont {C.}~\bibnamefont {Dallo}}, \bibinfo {author} {\bibfnamefont
  {A.}~\bibnamefont {Pomerene}}, \bibinfo {author} {\bibfnamefont
  {A.}~\bibnamefont {Starbuck}}, \bibinfo {author} {\bibfnamefont
  {D.}~\bibnamefont {Hood}}, \bibinfo {author} {\bibfnamefont {D.~C.}\
  \bibnamefont {Trotter}}, \bibinfo {author} {\bibfnamefont {T.}~\bibnamefont
  {Friedmann}}, \bibinfo {author} {\bibfnamefont {C.~T.}\ \bibnamefont
  {DeRose}},\ and\ \bibinfo {author} {\bibfnamefont {A.}~\bibnamefont
  {Lentine}},\ }\bibfield  {title} {\bibinfo {title} {A heterogeneously
  integrated silicon photonic/lithium niobate travelling wave electro-optic
  modulator},\ }\href {https://doi.org/10.1364/OE.28.001868} {\bibfield
  {journal} {\bibinfo  {journal} {Optics Express}\ }\textbf {\bibinfo {volume}
  {28}},\ \bibinfo {pages} {1868} (\bibinfo {year} {2020})}\BibitemShut
  {NoStop}%
\bibitem [{\citenamefont {Lee}\ \emph {et~al.}(2011)\citenamefont {Lee},
  \citenamefont {Kim}, \citenamefont {Kim}, \citenamefont {Lee}, \citenamefont
  {Lee},\ and\ \citenamefont {Steier}}]{leeHybridSiLiNbO3Microring2011}%
  \BibitemOpen
  \bibfield  {author} {\bibinfo {author} {\bibfnamefont {Y.~S.}\ \bibnamefont
  {Lee}}, \bibinfo {author} {\bibfnamefont {G.-D.}\ \bibnamefont {Kim}},
  \bibinfo {author} {\bibfnamefont {W.-J.}\ \bibnamefont {Kim}}, \bibinfo
  {author} {\bibfnamefont {S.-S.}\ \bibnamefont {Lee}}, \bibinfo {author}
  {\bibfnamefont {W.-G.}\ \bibnamefont {Lee}},\ and\ \bibinfo {author}
  {\bibfnamefont {W.~H.}\ \bibnamefont {Steier}},\ }\bibfield  {title}
  {\bibinfo {title} {Hybrid {{Si-LiNbO3}} microring electro-optically tunable
  resonators for active photonic devices},\ }\href
  {https://doi.org/10.1364/OL.36.001119} {\bibfield  {journal} {\bibinfo
  {journal} {Opt. Lett.}\ }\textbf {\bibinfo {volume} {36}},\ \bibinfo {pages}
  {1119} (\bibinfo {year} {2011})}\BibitemShut {NoStop}%
\bibitem [{\citenamefont {Rabiei}\ \emph {et~al.}(2013)\citenamefont {Rabiei},
  \citenamefont {Ma}, \citenamefont {Khan}, \citenamefont {Chiles},\ and\
  \citenamefont {Fathpour}}]{rabieiHeterogeneousLithiumNiobate2013}%
  \BibitemOpen
  \bibfield  {author} {\bibinfo {author} {\bibfnamefont {P.}~\bibnamefont
  {Rabiei}}, \bibinfo {author} {\bibfnamefont {J.}~\bibnamefont {Ma}}, \bibinfo
  {author} {\bibfnamefont {S.}~\bibnamefont {Khan}}, \bibinfo {author}
  {\bibfnamefont {J.}~\bibnamefont {Chiles}},\ and\ \bibinfo {author}
  {\bibfnamefont {S.}~\bibnamefont {Fathpour}},\ }\bibfield  {title} {\bibinfo
  {title} {Heterogeneous lithium niobate photonics on silicon substrates},\
  }\href {https://doi.org/10.1364/OE.21.025573} {\bibfield  {journal} {\bibinfo
   {journal} {Opt. Express}\ }\textbf {\bibinfo {volume} {21}},\ \bibinfo
  {pages} {25573} (\bibinfo {year} {2013})}\BibitemShut {NoStop}%
\bibitem [{\citenamefont {Bakish}\ \emph {et~al.}(2013)\citenamefont {Bakish},
  \citenamefont {Califa}, \citenamefont {Ilovitsh}, \citenamefont {Artel},
  \citenamefont {Winzer}, \citenamefont {Voigt}, \citenamefont {Zimmermann},
  \citenamefont {Shekel}, \citenamefont {Sukenik},\ and\ \citenamefont
  {Zadok}}]{bakishVoltageInducedPhaseShift2013}%
  \BibitemOpen
  \bibfield  {author} {\bibinfo {author} {\bibfnamefont {I.}~\bibnamefont
  {Bakish}}, \bibinfo {author} {\bibfnamefont {R.}~\bibnamefont {Califa}},
  \bibinfo {author} {\bibfnamefont {T.}~\bibnamefont {Ilovitsh}}, \bibinfo
  {author} {\bibfnamefont {V.}~\bibnamefont {Artel}}, \bibinfo {author}
  {\bibfnamefont {G.}~\bibnamefont {Winzer}}, \bibinfo {author} {\bibfnamefont
  {K.}~\bibnamefont {Voigt}}, \bibinfo {author} {\bibfnamefont
  {L.}~\bibnamefont {Zimmermann}}, \bibinfo {author} {\bibfnamefont
  {E.}~\bibnamefont {Shekel}}, \bibinfo {author} {\bibfnamefont
  {C.}~\bibnamefont {Sukenik}},\ and\ \bibinfo {author} {\bibfnamefont
  {A.}~\bibnamefont {Zadok}},\ }\bibfield  {title} {\bibinfo {title}
  {Voltage-{{Induced Phase Shift}} in a {{Hybrid LiNbO3-on-Silicon Mach-Zehnder
  Interferometer}}},\ }in\ \href {https://doi.org/10.1364/IPRSN.2013.IW4A.2}
  {\emph {\bibinfo {booktitle} {Advanced {{Photonics}} 2013}}}\ (\bibinfo
  {publisher} {OSA},\ \bibinfo {address} {Rio Grande, Puerto Rico},\ \bibinfo
  {year} {2013})\ p.\ \bibinfo {pages} {IW4A.2}\BibitemShut {NoStop}%
\bibitem [{\citenamefont {Mere}\ \emph {et~al.}(2022)\citenamefont {Mere},
  \citenamefont {Valdez}, \citenamefont {Wang},\ and\ \citenamefont
  {Mookherjea}}]{mereModularFabricationProcess2022}%
  \BibitemOpen
  \bibfield  {author} {\bibinfo {author} {\bibfnamefont {V.}~\bibnamefont
  {Mere}}, \bibinfo {author} {\bibfnamefont {F.}~\bibnamefont {Valdez}},
  \bibinfo {author} {\bibfnamefont {X.}~\bibnamefont {Wang}},\ and\ \bibinfo
  {author} {\bibfnamefont {S.}~\bibnamefont {Mookherjea}},\ }\bibfield  {title}
  {\bibinfo {title} {A modular fabrication process for thin-film lithium
  niobate modulators with silicon photonics},\ }\href
  {https://doi.org/10.1088/2515-7647/ac5e0b} {\bibfield  {journal} {\bibinfo
  {journal} {Journal of Physics: Photonics}\ }\textbf {\bibinfo {volume} {4}},\
  \bibinfo {pages} {024001} (\bibinfo {year} {2022})}\BibitemShut {NoStop}%
\bibitem [{\citenamefont {Mere}\ \emph {et~al.}(2023)\citenamefont {Mere},
  \citenamefont {Valdez},\ and\ \citenamefont
  {Mookherjea}}]{mereImprovedFabricationScalable2023}%
  \BibitemOpen
  \bibfield  {author} {\bibinfo {author} {\bibfnamefont {V.}~\bibnamefont
  {Mere}}, \bibinfo {author} {\bibfnamefont {F.}~\bibnamefont {Valdez}},\ and\
  \bibinfo {author} {\bibfnamefont {S.}~\bibnamefont {Mookherjea}},\ }\bibfield
   {title} {\bibinfo {title} {Improved fabrication of scalable hybrid silicon
  nitride--thin-film lithium niobate electro-optic modulators},\ }\href
  {https://doi.org/10.1364/JOSAB.481915} {\bibfield  {journal} {\bibinfo
  {journal} {Journal of the Optical Society of America B}\ }\textbf {\bibinfo
  {volume} {40}},\ \bibinfo {pages} {D5} (\bibinfo {year} {2023})}\BibitemShut
  {NoStop}%
\bibitem [{\citenamefont {Smith}\ \emph {et~al.}(2022)\citenamefont {Smith},
  \citenamefont {Jevtics}, \citenamefont {Guilhabert}, \citenamefont {Dawson},\
  and\ \citenamefont {Strain}}]{smithHybridIntegrationChipscale2022}%
  \BibitemOpen
  \bibfield  {author} {\bibinfo {author} {\bibfnamefont {J.~A.}\ \bibnamefont
  {Smith}}, \bibinfo {author} {\bibfnamefont {D.}~\bibnamefont {Jevtics}},
  \bibinfo {author} {\bibfnamefont {B.}~\bibnamefont {Guilhabert}}, \bibinfo
  {author} {\bibfnamefont {M.~D.}\ \bibnamefont {Dawson}},\ and\ \bibinfo
  {author} {\bibfnamefont {M.~J.}\ \bibnamefont {Strain}},\ }\bibfield  {title}
  {\bibinfo {title} {Hybrid integration of chipscale photonic devices using
  accurate transfer printing methods},\ }\href
  {https://doi.org/10.1063/5.0121567} {\bibfield  {journal} {\bibinfo
  {journal} {Applied Physics Reviews}\ }\textbf {\bibinfo {volume} {9}},\
  \bibinfo {pages} {041317} (\bibinfo {year} {2022})}\BibitemShut {NoStop}%
\bibitem [{\citenamefont {Vandekerckhove}\ \emph {et~al.}(2023)\citenamefont
  {Vandekerckhove}, \citenamefont {Vanackere}, \citenamefont {De~Witte},
  \citenamefont {Cuyvers}, \citenamefont {Reis}, \citenamefont {Billet},
  \citenamefont {Roelkens}, \citenamefont {Clemmen},\ and\ \citenamefont
  {Kuyken}}]{vandekerckhoveReliableMicrotransferPrinting2023}%
  \BibitemOpen
  \bibfield  {author} {\bibinfo {author} {\bibfnamefont {T.}~\bibnamefont
  {Vandekerckhove}}, \bibinfo {author} {\bibfnamefont {T.}~\bibnamefont
  {Vanackere}}, \bibinfo {author} {\bibfnamefont {J.}~\bibnamefont {De~Witte}},
  \bibinfo {author} {\bibfnamefont {S.}~\bibnamefont {Cuyvers}}, \bibinfo
  {author} {\bibfnamefont {L.}~\bibnamefont {Reis}}, \bibinfo {author}
  {\bibfnamefont {M.}~\bibnamefont {Billet}}, \bibinfo {author} {\bibfnamefont
  {G.}~\bibnamefont {Roelkens}}, \bibinfo {author} {\bibfnamefont
  {S.}~\bibnamefont {Clemmen}},\ and\ \bibinfo {author} {\bibfnamefont
  {B.}~\bibnamefont {Kuyken}},\ }\bibfield  {title} {\bibinfo {title} {Reliable
  micro-transfer printing method for heterogeneous integration of lithium
  niobate and semiconductor thin films},\ }\href
  {https://doi.org/10.1364/OME.494038} {\bibfield  {journal} {\bibinfo
  {journal} {Optical Materials Express}\ }\textbf {\bibinfo {volume} {13}},\
  \bibinfo {pages} {1984} (\bibinfo {year} {2023})}\BibitemShut {NoStop}%
\bibitem [{\citenamefont {Guarino}\ \emph {et~al.}(2007)\citenamefont
  {Guarino}, \citenamefont {Poberaj}, \citenamefont {Rezzonico}, \citenamefont
  {Degl'Innocenti},\ and\ \citenamefont
  {G{\"u}nter}}]{guarinoElectroOpticallyTunable2007}%
  \BibitemOpen
  \bibfield  {author} {\bibinfo {author} {\bibfnamefont {A.}~\bibnamefont
  {Guarino}}, \bibinfo {author} {\bibfnamefont {G.}~\bibnamefont {Poberaj}},
  \bibinfo {author} {\bibfnamefont {D.}~\bibnamefont {Rezzonico}}, \bibinfo
  {author} {\bibfnamefont {R.}~\bibnamefont {Degl'Innocenti}},\ and\ \bibinfo
  {author} {\bibfnamefont {P.}~\bibnamefont {G{\"u}nter}},\ }\bibfield  {title}
  {\bibinfo {title} {Electro--optically tunable microring resonators in lithium
  niobate},\ }\href@noop {} {\bibfield  {journal} {\bibinfo  {journal} {Nature
  Photonics}\ }\textbf {\bibinfo {volume} {1}},\ \bibinfo {pages} {407}
  (\bibinfo {year} {2007})}\BibitemShut {NoStop}%
\bibitem [{\citenamefont {Chen}\ \emph {et~al.}(2022)\citenamefont {Chen},
  \citenamefont {Chen}, \citenamefont {Gan}, \citenamefont {Ruan},
  \citenamefont {Wang}, \citenamefont {Huang}, \citenamefont {Lu},
  \citenamefont {Lau}, \citenamefont {Dai}, \citenamefont {Guo},\ and\
  \citenamefont {Liu}}]{chenHighPerformanceThinfilm2022}%
  \BibitemOpen
  \bibfield  {author} {\bibinfo {author} {\bibfnamefont {G.}~\bibnamefont
  {Chen}}, \bibinfo {author} {\bibfnamefont {K.}~\bibnamefont {Chen}}, \bibinfo
  {author} {\bibfnamefont {R.}~\bibnamefont {Gan}}, \bibinfo {author}
  {\bibfnamefont {Z.}~\bibnamefont {Ruan}}, \bibinfo {author} {\bibfnamefont
  {Z.}~\bibnamefont {Wang}}, \bibinfo {author} {\bibfnamefont {P.}~\bibnamefont
  {Huang}}, \bibinfo {author} {\bibfnamefont {C.}~\bibnamefont {Lu}}, \bibinfo
  {author} {\bibfnamefont {A.~P.~T.}\ \bibnamefont {Lau}}, \bibinfo {author}
  {\bibfnamefont {D.}~\bibnamefont {Dai}}, \bibinfo {author} {\bibfnamefont
  {C.}~\bibnamefont {Guo}},\ and\ \bibinfo {author} {\bibfnamefont
  {L.}~\bibnamefont {Liu}},\ }\bibfield  {title} {\bibinfo {title} {High
  performance thin-film lithium niobate modulator on a silicon substrate using
  periodic capacitively loaded traveling-wave electrode},\ }\href
  {https://doi.org/10.1063/5.0077232} {\bibfield  {journal} {\bibinfo
  {journal} {APL Photonics}\ }\textbf {\bibinfo {volume} {7}},\ \bibinfo
  {pages} {026103} (\bibinfo {year} {2022})}\BibitemShut {NoStop}%
\bibitem [{\citenamefont {Della~Torre}\ \emph {et~al.}(2025)\citenamefont
  {Della~Torre}, \citenamefont {Dubois}, \citenamefont {Zarebidaki},
  \citenamefont {Volpini}, \citenamefont {Leo}, \citenamefont {Mettraux},
  \citenamefont {Manzoor}, \citenamefont {Prieto}, \citenamefont {Grassani},
  \citenamefont {Dubochet}, \citenamefont {Despont},\ and\ \citenamefont
  {Sattari}}]{dellatorreFoldedElectroopticalModulators2025}%
  \BibitemOpen
  \bibfield  {author} {\bibinfo {author} {\bibfnamefont {A.}~\bibnamefont
  {Della~Torre}}, \bibinfo {author} {\bibfnamefont {F.}~\bibnamefont {Dubois}},
  \bibinfo {author} {\bibfnamefont {H.}~\bibnamefont {Zarebidaki}}, \bibinfo
  {author} {\bibfnamefont {A.}~\bibnamefont {Volpini}}, \bibinfo {author}
  {\bibfnamefont {J.}~\bibnamefont {Leo}}, \bibinfo {author} {\bibfnamefont
  {A.}~\bibnamefont {Mettraux}}, \bibinfo {author} {\bibfnamefont
  {A.}~\bibnamefont {Manzoor}}, \bibinfo {author} {\bibfnamefont
  {I.}~\bibnamefont {Prieto}}, \bibinfo {author} {\bibfnamefont
  {D.}~\bibnamefont {Grassani}}, \bibinfo {author} {\bibfnamefont
  {O.}~\bibnamefont {Dubochet}}, \bibinfo {author} {\bibfnamefont
  {M.}~\bibnamefont {Despont}},\ and\ \bibinfo {author} {\bibfnamefont
  {H.}~\bibnamefont {Sattari}},\ }\bibfield  {title} {\bibinfo {title} {Folded
  electro-optical modulators operating at {{CMOS}} voltage level in a thin-film
  lithium niobate foundry process},\ }\href {https://doi.org/10.1364/OE.548003}
  {\bibfield  {journal} {\bibinfo  {journal} {Optics Express}\ }\textbf
  {\bibinfo {volume} {33}},\ \bibinfo {pages} {6747} (\bibinfo {year}
  {2025})}\BibitemShut {NoStop}%
\bibitem [{\citenamefont {Mookherjea}\ \emph {et~al.}(2023)\citenamefont
  {Mookherjea}, \citenamefont {Mere},\ and\ \citenamefont
  {Valdez}}]{mookherjeaThinfilmLithiumNiobate2023}%
  \BibitemOpen
  \bibfield  {author} {\bibinfo {author} {\bibfnamefont {S.}~\bibnamefont
  {Mookherjea}}, \bibinfo {author} {\bibfnamefont {V.}~\bibnamefont {Mere}},\
  and\ \bibinfo {author} {\bibfnamefont {F.}~\bibnamefont {Valdez}},\
  }\bibfield  {title} {\bibinfo {title} {Thin-film lithium niobate
  electro-optic modulators: {{To}} etch or not to etch},\ }\href
  {https://doi.org/10.1063/5.0142232} {\bibfield  {journal} {\bibinfo
  {journal} {Applied Physics Letters}\ }\textbf {\bibinfo {volume} {122}},\
  \bibinfo {pages} {120501} (\bibinfo {year} {2023})}\BibitemShut {NoStop}%
\bibitem [{\citenamefont {Wang}\ \emph {et~al.}(2019)\citenamefont {Wang},
  \citenamefont {Weigel}, \citenamefont {Zhao}, \citenamefont {Ruesing},\ and\
  \citenamefont {Mookherjea}}]{wangAchievingBeyond100GHzLargesignal2019a}%
  \BibitemOpen
  \bibfield  {author} {\bibinfo {author} {\bibfnamefont {X.}~\bibnamefont
  {Wang}}, \bibinfo {author} {\bibfnamefont {P.~O.}\ \bibnamefont {Weigel}},
  \bibinfo {author} {\bibfnamefont {J.}~\bibnamefont {Zhao}}, \bibinfo {author}
  {\bibfnamefont {M.}~\bibnamefont {Ruesing}},\ and\ \bibinfo {author}
  {\bibfnamefont {S.}~\bibnamefont {Mookherjea}},\ }\bibfield  {title}
  {\bibinfo {title} {Achieving beyond-100-{{GHz}} large-signal modulation
  bandwidth in hybrid silicon photonics {{Mach Zehnder}} modulators using thin
  film lithium niobate},\ }\href {https://doi.org/10.1063/1.5115243} {\bibfield
   {journal} {\bibinfo  {journal} {APL Photonics}\ }\textbf {\bibinfo {volume}
  {4}},\ \bibinfo {pages} {096101} (\bibinfo {year} {2019})}\BibitemShut
  {NoStop}%
\bibitem [{\citenamefont {Valdez}\ \emph
  {et~al.}(2022{\natexlab{a}})\citenamefont {Valdez}, \citenamefont {Mere},
  \citenamefont {Wang}, \citenamefont {Boynton}, \citenamefont {Friedmann},
  \citenamefont {Arterburn}, \citenamefont {Dallo}, \citenamefont {Pomerene},
  \citenamefont {Starbuck}, \citenamefont {Trotter}, \citenamefont {Lentine},\
  and\ \citenamefont {Mookherjea}}]{valdez110GHz1102022a}%
  \BibitemOpen
  \bibfield  {author} {\bibinfo {author} {\bibfnamefont {F.}~\bibnamefont
  {Valdez}}, \bibinfo {author} {\bibfnamefont {V.}~\bibnamefont {Mere}},
  \bibinfo {author} {\bibfnamefont {X.}~\bibnamefont {Wang}}, \bibinfo {author}
  {\bibfnamefont {N.}~\bibnamefont {Boynton}}, \bibinfo {author} {\bibfnamefont
  {T.~A.}\ \bibnamefont {Friedmann}}, \bibinfo {author} {\bibfnamefont
  {S.}~\bibnamefont {Arterburn}}, \bibinfo {author} {\bibfnamefont
  {C.}~\bibnamefont {Dallo}}, \bibinfo {author} {\bibfnamefont {A.~T.}\
  \bibnamefont {Pomerene}}, \bibinfo {author} {\bibfnamefont {A.~L.}\
  \bibnamefont {Starbuck}}, \bibinfo {author} {\bibfnamefont {D.~C.}\
  \bibnamefont {Trotter}}, \bibinfo {author} {\bibfnamefont {A.~L.}\
  \bibnamefont {Lentine}},\ and\ \bibinfo {author} {\bibfnamefont
  {S.}~\bibnamefont {Mookherjea}},\ }\bibfield  {title} {\bibinfo {title} {110
  {{GHz}}, 110 {{mW}} hybrid silicon-lithium niobate {{Mach-Zehnder}}
  modulator},\ }\href {https://doi.org/10.1038/s41598-022-23403-6} {\bibfield
  {journal} {\bibinfo  {journal} {Scientific Reports}\ }\textbf {\bibinfo
  {volume} {12}},\ \bibinfo {pages} {18611} (\bibinfo {year}
  {2022}{\natexlab{a}})}\BibitemShut {NoStop}%
\bibitem [{\citenamefont {Valdez}\ \emph
  {et~al.}(2023{\natexlab{a}})\citenamefont {Valdez}, \citenamefont {Mere},
  \citenamefont {Wang},\ and\ \citenamefont
  {Mookherjea}}]{valdezIntegratedCbandSiliconlithium2023}%
  \BibitemOpen
  \bibfield  {author} {\bibinfo {author} {\bibfnamefont {F.}~\bibnamefont
  {Valdez}}, \bibinfo {author} {\bibfnamefont {V.}~\bibnamefont {Mere}},
  \bibinfo {author} {\bibfnamefont {X.}~\bibnamefont {Wang}},\ and\ \bibinfo
  {author} {\bibfnamefont {S.}~\bibnamefont {Mookherjea}},\ }\bibfield  {title}
  {\bibinfo {title} {Integrated {{O-}} and {{C-band}} silicon-lithium niobate
  {{Mach-Zehnder}} modulators with 100 {{GHz}} bandwidth, low voltage, and low
  loss},\ }\href {https://doi.org/10.1364/OE.480519} {\bibfield  {journal}
  {\bibinfo  {journal} {Optics Express}\ }\textbf {\bibinfo {volume} {31}},\
  \bibinfo {pages} {5273} (\bibinfo {year} {2023}{\natexlab{a}})}\BibitemShut
  {NoStop}%
\bibitem [{\citenamefont {Valdez}\ \emph
  {et~al.}(2023{\natexlab{b}})\citenamefont {Valdez}, \citenamefont {Mere},\
  and\ \citenamefont {Mookherjea}}]{valdez100GHzBandwidth2023}%
  \BibitemOpen
  \bibfield  {author} {\bibinfo {author} {\bibfnamefont {F.}~\bibnamefont
  {Valdez}}, \bibinfo {author} {\bibfnamefont {V.}~\bibnamefont {Mere}},\ and\
  \bibinfo {author} {\bibfnamefont {S.}~\bibnamefont {Mookherjea}},\ }\bibfield
   {title} {\bibinfo {title} {100 {{GHz Bandwidth}}, 1 {{Volt Near-infrared
  Electro-optic Mach-Zehnder Modulator}}},\ }\href
  {https://doi.org/10.1364/OPTICA.484549} {\bibfield  {journal} {\bibinfo
  {journal} {Optica}\ }\textbf {\bibinfo {volume} {10}},\ \bibinfo {pages}
  {578} (\bibinfo {year} {2023}{\natexlab{b}})},\ \Eprint
  {https://arxiv.org/abs/2211.13348} {arXiv:2211.13348 [physics]} \BibitemShut
  {NoStop}%
\bibitem [{\citenamefont {Xue}\ \emph {et~al.}(2022)\citenamefont {Xue},
  \citenamefont {Gan}, \citenamefont {Chen}, \citenamefont {Chen},
  \citenamefont {Ruan}, \citenamefont {Zhang}, \citenamefont {Liu},
  \citenamefont {Dai}, \citenamefont {Guo},\ and\ \citenamefont
  {Liu}}]{xueBreakingBandwidthLimit2022}%
  \BibitemOpen
  \bibfield  {author} {\bibinfo {author} {\bibfnamefont {Y.}~\bibnamefont
  {Xue}}, \bibinfo {author} {\bibfnamefont {R.}~\bibnamefont {Gan}}, \bibinfo
  {author} {\bibfnamefont {K.}~\bibnamefont {Chen}}, \bibinfo {author}
  {\bibfnamefont {G.}~\bibnamefont {Chen}}, \bibinfo {author} {\bibfnamefont
  {Z.}~\bibnamefont {Ruan}}, \bibinfo {author} {\bibfnamefont {J.}~\bibnamefont
  {Zhang}}, \bibinfo {author} {\bibfnamefont {J.}~\bibnamefont {Liu}}, \bibinfo
  {author} {\bibfnamefont {D.}~\bibnamefont {Dai}}, \bibinfo {author}
  {\bibfnamefont {C.}~\bibnamefont {Guo}},\ and\ \bibinfo {author}
  {\bibfnamefont {L.}~\bibnamefont {Liu}},\ }\bibfield  {title} {\bibinfo
  {title} {Breaking the bandwidth limit of a high-quality-factor ring modulator
  based on thin-film lithium niobate},\ }\href
  {https://doi.org/10.1364/OPTICA.470596} {\bibfield  {journal} {\bibinfo
  {journal} {Optica}\ }\textbf {\bibinfo {volume} {9}},\ \bibinfo {pages}
  {1131} (\bibinfo {year} {2022})}\BibitemShut {NoStop}%
\bibitem [{\citenamefont {Pan}\ \emph {et~al.}(2022)\citenamefont {Pan},
  \citenamefont {He}, \citenamefont {Xu}, \citenamefont {Lin}, \citenamefont
  {Lin}, \citenamefont {Ke}, \citenamefont {Liu}, \citenamefont {Lin},
  \citenamefont {Zhu}, \citenamefont {Gao}, \citenamefont {Li}, \citenamefont
  {Liu}, \citenamefont {Liu}, \citenamefont {Yu},\ and\ \citenamefont
  {Cai}}]{panCompactSubstrateremovedThinfilm2022}%
  \BibitemOpen
  \bibfield  {author} {\bibinfo {author} {\bibfnamefont {Y.}~\bibnamefont
  {Pan}}, \bibinfo {author} {\bibfnamefont {M.}~\bibnamefont {He}}, \bibinfo
  {author} {\bibfnamefont {M.}~\bibnamefont {Xu}}, \bibinfo {author}
  {\bibfnamefont {Z.}~\bibnamefont {Lin}}, \bibinfo {author} {\bibfnamefont
  {Y.}~\bibnamefont {Lin}}, \bibinfo {author} {\bibfnamefont {W.}~\bibnamefont
  {Ke}}, \bibinfo {author} {\bibfnamefont {J.}~\bibnamefont {Liu}}, \bibinfo
  {author} {\bibfnamefont {Z.}~\bibnamefont {Lin}}, \bibinfo {author}
  {\bibfnamefont {Y.}~\bibnamefont {Zhu}}, \bibinfo {author} {\bibfnamefont
  {S.}~\bibnamefont {Gao}}, \bibinfo {author} {\bibfnamefont {H.}~\bibnamefont
  {Li}}, \bibinfo {author} {\bibfnamefont {X.}~\bibnamefont {Liu}}, \bibinfo
  {author} {\bibfnamefont {C.}~\bibnamefont {Liu}}, \bibinfo {author}
  {\bibfnamefont {S.}~\bibnamefont {Yu}},\ and\ \bibinfo {author}
  {\bibfnamefont {X.}~\bibnamefont {Cai}},\ }\bibfield  {title} {\bibinfo
  {title} {Compact substrate-removed thin-film lithium niobate electro-optic
  modulator featuring polarization-insensitive operation},\ }\href
  {https://doi.org/10.1364/OL.454277} {\bibfield  {journal} {\bibinfo
  {journal} {Optics Letters}\ }\textbf {\bibinfo {volume} {47}},\ \bibinfo
  {pages} {1818} (\bibinfo {year} {2022})}\BibitemShut {NoStop}%
\bibitem [{\citenamefont {Xu}\ \emph {et~al.}(2022)\citenamefont {Xu},
  \citenamefont {Gao}, \citenamefont {Tan},\ and\ \citenamefont
  {Cai}}]{xuCMOSlevelvoltageSubstrateremovedThinfilm2022}%
  \BibitemOpen
  \bibfield  {author} {\bibinfo {author} {\bibfnamefont {M.}~\bibnamefont
  {Xu}}, \bibinfo {author} {\bibfnamefont {S.}~\bibnamefont {Gao}}, \bibinfo
  {author} {\bibfnamefont {H.}~\bibnamefont {Tan}},\ and\ \bibinfo {author}
  {\bibfnamefont {X.}~\bibnamefont {Cai}},\ }\bibfield  {title} {\bibinfo
  {title} {{{CMOS-level-voltage Substrate-removed Thin-film Lithium Niobate
  Modulator}}},\ }in\ \href {https://doi.org/10.1364/OFC.2022.Th1J.3} {\emph
  {\bibinfo {booktitle} {Optical {{Fiber Communication Conference}} ({{OFC}})
  2022}}}\ (\bibinfo  {publisher} {Optica Publishing Group},\ \bibinfo
  {address} {San Diego, California},\ \bibinfo {year} {2022})\ p.\ \bibinfo
  {pages} {Th1J.3}\BibitemShut {NoStop}%
\bibitem [{\citenamefont {Sakamoto}\ \emph {et~al.}(1995)\citenamefont
  {Sakamoto}, \citenamefont {Spickermann},\ and\ \citenamefont
  {Dagli}}]{sakamotoNarrowGapCoplanar1995}%
  \BibitemOpen
  \bibfield  {author} {\bibinfo {author} {\bibfnamefont {S.}~\bibnamefont
  {Sakamoto}}, \bibinfo {author} {\bibfnamefont {R.}~\bibnamefont
  {Spickermann}},\ and\ \bibinfo {author} {\bibfnamefont {N.}~\bibnamefont
  {Dagli}},\ }\bibfield  {title} {\bibinfo {title} {Narrow gap coplanar slow
  wave electrode for travelling wave electro-optic modulators},\ }\href
  {https://doi.org/10.1049/el:19950779} {\bibfield  {journal} {\bibinfo
  {journal} {Electronics Letters}\ }\textbf {\bibinfo {volume} {31}},\ \bibinfo
  {pages} {1183} (\bibinfo {year} {1995})}\BibitemShut {NoStop}%
\bibitem [{\citenamefont {Rosa}\ \emph {et~al.}(2018)\citenamefont {Rosa},
  \citenamefont {Verstuyft}, \citenamefont {Brimont}, \citenamefont
  {Thourhout},\ and\ \citenamefont
  {Sanchis}}]{rosaMicrowaveIndexEngineering2018}%
  \BibitemOpen
  \bibfield  {author} {\bibinfo {author} {\bibfnamefont {{\'A}.}~\bibnamefont
  {Rosa}}, \bibinfo {author} {\bibfnamefont {S.}~\bibnamefont {Verstuyft}},
  \bibinfo {author} {\bibfnamefont {A.}~\bibnamefont {Brimont}}, \bibinfo
  {author} {\bibfnamefont {D.~V.}\ \bibnamefont {Thourhout}},\ and\ \bibinfo
  {author} {\bibfnamefont {P.}~\bibnamefont {Sanchis}},\ }\bibfield  {title}
  {\bibinfo {title} {Microwave index engineering for slow-wave coplanar
  waveguides},\ }\href {https://doi.org/10.1038/s41598-018-24030-w} {\bibfield
  {journal} {\bibinfo  {journal} {Scientific Reports}\ }\textbf {\bibinfo
  {volume} {8}},\ \bibinfo {pages} {5672} (\bibinfo {year} {2018})}\BibitemShut
  {NoStop}%
\bibitem [{\citenamefont {Xu}\ \emph {et~al.}(2018)\citenamefont {Xu},
  \citenamefont {Wang}, \citenamefont {Tian}, \citenamefont {Wu}, \citenamefont
  {Wang},\ and\ \citenamefont
  {Zhang}}]{xuGlassonLiNbO3HeterostructureFormed2018}%
  \BibitemOpen
  \bibfield  {author} {\bibinfo {author} {\bibfnamefont {J.}~\bibnamefont
  {Xu}}, \bibinfo {author} {\bibfnamefont {C.}~\bibnamefont {Wang}}, \bibinfo
  {author} {\bibfnamefont {Y.}~\bibnamefont {Tian}}, \bibinfo {author}
  {\bibfnamefont {B.}~\bibnamefont {Wu}}, \bibinfo {author} {\bibfnamefont
  {S.}~\bibnamefont {Wang}},\ and\ \bibinfo {author} {\bibfnamefont
  {H.}~\bibnamefont {Zhang}},\ }\bibfield  {title} {\bibinfo {title}
  {Glass-on-{{LiNbO3}} heterostructure formed via a two-step plasma activated
  low-temperature direct bonding method},\ }\href
  {https://doi.org/10.1016/j.apsusc.2018.08.031} {\bibfield  {journal}
  {\bibinfo  {journal} {Applied Surface Science}\ }\textbf {\bibinfo {volume}
  {459}},\ \bibinfo {pages} {621} (\bibinfo {year} {2018})}\BibitemShut
  {NoStop}%
\bibitem [{\citenamefont {Valdez}\ \emph
  {et~al.}(2023{\natexlab{c}})\citenamefont {Valdez}, \citenamefont {Mere},
  \citenamefont {Boynton}, \citenamefont {Friedmann}, \citenamefont
  {Arterburn}, \citenamefont {Dallo}, \citenamefont {Pomerene}, \citenamefont
  {Starbuck}, \citenamefont {Trotter}, \citenamefont {Lentine}, \citenamefont
  {Kodigala},\ and\ \citenamefont
  {Mookherjea}}]{valdezBuriedElectrodeHybridBonded2023}%
  \BibitemOpen
  \bibfield  {author} {\bibinfo {author} {\bibfnamefont {F.}~\bibnamefont
  {Valdez}}, \bibinfo {author} {\bibfnamefont {V.}~\bibnamefont {Mere}},
  \bibinfo {author} {\bibfnamefont {N.}~\bibnamefont {Boynton}}, \bibinfo
  {author} {\bibfnamefont {T.~A.}\ \bibnamefont {Friedmann}}, \bibinfo {author}
  {\bibfnamefont {S.}~\bibnamefont {Arterburn}}, \bibinfo {author}
  {\bibfnamefont {C.}~\bibnamefont {Dallo}}, \bibinfo {author} {\bibfnamefont
  {A.~T.}\ \bibnamefont {Pomerene}}, \bibinfo {author} {\bibfnamefont {A.~L.}\
  \bibnamefont {Starbuck}}, \bibinfo {author} {\bibfnamefont {D.~C.}\
  \bibnamefont {Trotter}}, \bibinfo {author} {\bibfnamefont {A.~L.}\
  \bibnamefont {Lentine}}, \bibinfo {author} {\bibfnamefont {A.}~\bibnamefont
  {Kodigala}},\ and\ \bibinfo {author} {\bibfnamefont {S.}~\bibnamefont
  {Mookherjea}},\ }\bibfield  {title} {\bibinfo {title} {Buried-{{Electrode
  Hybrid Bonded Thin-Film Lithium Niobate Electro-Optic Mach-Zehnder
  Modulators}}},\ }\href {https://doi.org/10.1109/LPT.2023.3268605} {\bibfield
  {journal} {\bibinfo  {journal} {IEEE Photonics Technology Letters}\ }\textbf
  {\bibinfo {volume} {35}},\ \bibinfo {pages} {633} (\bibinfo {year}
  {2023}{\natexlab{c}})}\BibitemShut {NoStop}%
\bibitem [{\citenamefont {Shi}\ \emph {et~al.}(2003)\citenamefont {Shi},
  \citenamefont {Yan},\ and\ \citenamefont
  {Willner}}]{shiHighspeedElectroopticModulator2003}%
  \BibitemOpen
  \bibfield  {author} {\bibinfo {author} {\bibfnamefont {Y.}~\bibnamefont
  {Shi}}, \bibinfo {author} {\bibfnamefont {L.}~\bibnamefont {Yan}},\ and\
  \bibinfo {author} {\bibfnamefont {A.~E.}\ \bibnamefont {Willner}},\
  }\bibfield  {title} {\bibinfo {title} {High-speed electrooptic modulator
  characterization using optical spectrum analysis},\ }\href
  {https://doi.org/10.1109/JLT.2003.818162} {\bibfield  {journal} {\bibinfo
  {journal} {Journal of Lightwave Technology}\ }\textbf {\bibinfo {volume}
  {21}},\ \bibinfo {pages} {2358} (\bibinfo {year} {2003})}\BibitemShut
  {NoStop}%
\bibitem [{\citenamefont {Lee}\ \emph {et~al.}(2002)\citenamefont {Lee},
  \citenamefont {Katz}, \citenamefont {Erben}, \citenamefont {Gill},
  \citenamefont {Gopalan}, \citenamefont {Heber},\ and\ \citenamefont
  {McGee}}]{leeBroadbandModulationLight2002}%
  \BibitemOpen
  \bibfield  {author} {\bibinfo {author} {\bibfnamefont {M.}~\bibnamefont
  {Lee}}, \bibinfo {author} {\bibfnamefont {H.~E.}\ \bibnamefont {Katz}},
  \bibinfo {author} {\bibfnamefont {C.}~\bibnamefont {Erben}}, \bibinfo
  {author} {\bibfnamefont {D.~M.}\ \bibnamefont {Gill}}, \bibinfo {author}
  {\bibfnamefont {P.}~\bibnamefont {Gopalan}}, \bibinfo {author} {\bibfnamefont
  {J.~D.}\ \bibnamefont {Heber}},\ and\ \bibinfo {author} {\bibfnamefont
  {D.~J.}\ \bibnamefont {McGee}},\ }\bibfield  {title} {\bibinfo {title}
  {Broadband {{Modulation}} of {{Light}} by {{Using}} an {{Electro-Optic
  Polymer}}},\ }\href {https://doi.org/10.1126/science.1077446} {\bibfield
  {journal} {\bibinfo  {journal} {Science}\ }\textbf {\bibinfo {volume}
  {298}},\ \bibinfo {pages} {1401} (\bibinfo {year} {2002})}\BibitemShut
  {NoStop}%
\bibitem [{\citenamefont {Haffner}\ \emph {et~al.}(2018)\citenamefont
  {Haffner}, \citenamefont {Chelladurai}, \citenamefont {Fedoryshyn},
  \citenamefont {Josten}, \citenamefont {Baeuerle}, \citenamefont {Heni},
  \citenamefont {Watanabe}, \citenamefont {Cui}, \citenamefont {Cheng},
  \citenamefont {Saha}, \citenamefont {Elder}, \citenamefont {Dalton},
  \citenamefont {Boltasseva}, \citenamefont {Shalaev}, \citenamefont {Kinsey},\
  and\ \citenamefont
  {Leuthold}}]{haffnerLowlossPlasmonassistedElectrooptic2018}%
  \BibitemOpen
  \bibfield  {author} {\bibinfo {author} {\bibfnamefont {C.}~\bibnamefont
  {Haffner}}, \bibinfo {author} {\bibfnamefont {D.}~\bibnamefont
  {Chelladurai}}, \bibinfo {author} {\bibfnamefont {Y.}~\bibnamefont
  {Fedoryshyn}}, \bibinfo {author} {\bibfnamefont {A.}~\bibnamefont {Josten}},
  \bibinfo {author} {\bibfnamefont {B.}~\bibnamefont {Baeuerle}}, \bibinfo
  {author} {\bibfnamefont {W.}~\bibnamefont {Heni}}, \bibinfo {author}
  {\bibfnamefont {T.}~\bibnamefont {Watanabe}}, \bibinfo {author}
  {\bibfnamefont {T.}~\bibnamefont {Cui}}, \bibinfo {author} {\bibfnamefont
  {B.}~\bibnamefont {Cheng}}, \bibinfo {author} {\bibfnamefont
  {S.}~\bibnamefont {Saha}}, \bibinfo {author} {\bibfnamefont {D.~L.}\
  \bibnamefont {Elder}}, \bibinfo {author} {\bibfnamefont {{\relax Larry}.~R.}\
  \bibnamefont {Dalton}}, \bibinfo {author} {\bibfnamefont {A.}~\bibnamefont
  {Boltasseva}}, \bibinfo {author} {\bibfnamefont {V.~M.}\ \bibnamefont
  {Shalaev}}, \bibinfo {author} {\bibfnamefont {N.}~\bibnamefont {Kinsey}},\
  and\ \bibinfo {author} {\bibfnamefont {J.}~\bibnamefont {Leuthold}},\
  }\bibfield  {title} {\bibinfo {title} {Low-loss plasmon-assisted
  electro-optic modulator},\ }\href {https://doi.org/10.1038/s41586-018-0031-4}
  {\bibfield  {journal} {\bibinfo  {journal} {Nature}\ }\textbf {\bibinfo
  {volume} {556}},\ \bibinfo {pages} {483} (\bibinfo {year}
  {2018})}\BibitemShut {NoStop}%
\bibitem [{\citenamefont {Renaud}\ \emph {et~al.}(2023)\citenamefont {Renaud},
  \citenamefont {Assumpcao}, \citenamefont {Joe}, \citenamefont
  {{Shams-Ansari}}, \citenamefont {Zhu}, \citenamefont {Hu}, \citenamefont
  {Sinclair},\ and\ \citenamefont {Loncar}}]{renaudSub1VoltHighBandwidth2023}%
  \BibitemOpen
  \bibfield  {author} {\bibinfo {author} {\bibfnamefont {D.}~\bibnamefont
  {Renaud}}, \bibinfo {author} {\bibfnamefont {D.~R.}\ \bibnamefont
  {Assumpcao}}, \bibinfo {author} {\bibfnamefont {G.}~\bibnamefont {Joe}},
  \bibinfo {author} {\bibfnamefont {A.}~\bibnamefont {{Shams-Ansari}}},
  \bibinfo {author} {\bibfnamefont {D.}~\bibnamefont {Zhu}}, \bibinfo {author}
  {\bibfnamefont {Y.}~\bibnamefont {Hu}}, \bibinfo {author} {\bibfnamefont
  {N.}~\bibnamefont {Sinclair}},\ and\ \bibinfo {author} {\bibfnamefont
  {M.}~\bibnamefont {Loncar}},\ }\bibfield  {title} {\bibinfo {title} {Sub-1
  {{Volt}} and high-bandwidth visible to near-infrared electro-optic
  modulators},\ }\href {https://doi.org/10.1038/s41467-023-36870-w} {\bibfield
  {journal} {\bibinfo  {journal} {Nature Communications}\ }\textbf {\bibinfo
  {volume} {14}},\ \bibinfo {pages} {1496} (\bibinfo {year}
  {2023})}\BibitemShut {NoStop}%
\bibitem [{\citenamefont
  {Ghione}(2009)}]{ghioneSemiconductorDevicesHighspeed2009}%
  \BibitemOpen
  \bibfield  {author} {\bibinfo {author} {\bibfnamefont {G.}~\bibnamefont
  {Ghione}},\ }\href@noop {} {\emph {\bibinfo {title} {Semiconductor Devices
  for High-Speed Optoelectronics}}}\ (\bibinfo  {publisher} {Cambridge
  University},\ \bibinfo {year} {2009})\BibitemShut {NoStop}%
\bibitem [{\citenamefont {Pozar}(2012)}]{pozarMicrowaveEngineering2012}%
  \BibitemOpen
  \bibfield  {author} {\bibinfo {author} {\bibfnamefont {D.~M.}\ \bibnamefont
  {Pozar}},\ }\href@noop {} {\emph {\bibinfo {title} {Microwave
  {{Engineering}}}}},\ \bibinfo {edition} {4th}\ ed.\ (\bibinfo  {publisher}
  {Wiley},\ \bibinfo {year} {2012})\BibitemShut {NoStop}%
\bibitem [{\citenamefont {Weigel}\ \emph {et~al.}(2021)\citenamefont {Weigel},
  \citenamefont {Valdez}, \citenamefont {Zhao}, \citenamefont {Li},\ and\
  \citenamefont {Mookherjea}}]{weigelDesignHighbandwidthLowvoltage2021}%
  \BibitemOpen
  \bibfield  {author} {\bibinfo {author} {\bibfnamefont {P.~O.}\ \bibnamefont
  {Weigel}}, \bibinfo {author} {\bibfnamefont {F.}~\bibnamefont {Valdez}},
  \bibinfo {author} {\bibfnamefont {J.}~\bibnamefont {Zhao}}, \bibinfo {author}
  {\bibfnamefont {H.}~\bibnamefont {Li}},\ and\ \bibinfo {author}
  {\bibfnamefont {S.}~\bibnamefont {Mookherjea}},\ }\bibfield  {title}
  {\bibinfo {title} {Design of high-bandwidth, low-voltage and low-loss hybrid
  lithium niobate electro-optic modulators},\ }\href
  {https://doi.org/10.1088/2515-7647/abc17e} {\bibfield  {journal} {\bibinfo
  {journal} {Journal of Physics: Photonics}\ }\textbf {\bibinfo {volume} {3}},\
  \bibinfo {pages} {012001} (\bibinfo {year} {2021})}\BibitemShut {NoStop}%
\bibitem [{\citenamefont {Valdez}\ \emph
  {et~al.}(2022{\natexlab{b}})\citenamefont {Valdez}, \citenamefont {Mere},
  \citenamefont {Wang}, \citenamefont {Boynton}, \citenamefont {Friedmann},
  \citenamefont {Arterburn}, \citenamefont {Dallo}, \citenamefont {Pomerene},
  \citenamefont {Starbuck}, \citenamefont {Trotter}, \citenamefont {Lentine},\
  and\ \citenamefont {Mookherjea}}]{valdez110GHz1102022}%
  \BibitemOpen
  \bibfield  {author} {\bibinfo {author} {\bibfnamefont {F.}~\bibnamefont
  {Valdez}}, \bibinfo {author} {\bibfnamefont {V.}~\bibnamefont {Mere}},
  \bibinfo {author} {\bibfnamefont {X.}~\bibnamefont {Wang}}, \bibinfo {author}
  {\bibfnamefont {N.}~\bibnamefont {Boynton}}, \bibinfo {author} {\bibfnamefont
  {T.~A.}\ \bibnamefont {Friedmann}}, \bibinfo {author} {\bibfnamefont
  {S.}~\bibnamefont {Arterburn}}, \bibinfo {author} {\bibfnamefont
  {C.}~\bibnamefont {Dallo}}, \bibinfo {author} {\bibfnamefont {A.~T.}\
  \bibnamefont {Pomerene}}, \bibinfo {author} {\bibfnamefont {A.~L.}\
  \bibnamefont {Starbuck}}, \bibinfo {author} {\bibfnamefont {D.~C.}\
  \bibnamefont {Trotter}}, \bibinfo {author} {\bibfnamefont {A.~L.}\
  \bibnamefont {Lentine}},\ and\ \bibinfo {author} {\bibfnamefont
  {S.}~\bibnamefont {Mookherjea}},\ }\bibfield  {title} {\bibinfo {title} {110
  {{GHz}}, 110 {{mW}} hybrid silicon-lithium niobate {{Mach-Zehnder}}
  modulator},\ }\href {https://doi.org/10.1038/s41598-022-23403-6} {\bibfield
  {journal} {\bibinfo  {journal} {Scientific Reports}\ }\textbf {\bibinfo
  {volume} {12}},\ \bibinfo {pages} {18611} (\bibinfo {year}
  {2022}{\natexlab{b}})}\BibitemShut {NoStop}%
\bibitem [{\citenamefont {Churaev}\ \emph {et~al.}(2023)\citenamefont
  {Churaev}, \citenamefont {Wang}, \citenamefont {Riedhauser}, \citenamefont
  {Snigirev}, \citenamefont {Bl{\'e}sin}, \citenamefont {M{\"o}hl},
  \citenamefont {Anderson}, \citenamefont {Siddharth}, \citenamefont {Popoff},
  \citenamefont {Drechsler}, \citenamefont {Caimi}, \citenamefont {H{\"o}nl},
  \citenamefont {Riemensberger}, \citenamefont {Liu}, \citenamefont {Seidler},\
  and\ \citenamefont
  {Kippenberg}}]{churaevHeterogeneouslyIntegratedLithium2023}%
  \BibitemOpen
  \bibfield  {author} {\bibinfo {author} {\bibfnamefont {M.}~\bibnamefont
  {Churaev}}, \bibinfo {author} {\bibfnamefont {R.~N.}\ \bibnamefont {Wang}},
  \bibinfo {author} {\bibfnamefont {A.}~\bibnamefont {Riedhauser}}, \bibinfo
  {author} {\bibfnamefont {V.}~\bibnamefont {Snigirev}}, \bibinfo {author}
  {\bibfnamefont {T.}~\bibnamefont {Bl{\'e}sin}}, \bibinfo {author}
  {\bibfnamefont {C.}~\bibnamefont {M{\"o}hl}}, \bibinfo {author}
  {\bibfnamefont {M.~H.}\ \bibnamefont {Anderson}}, \bibinfo {author}
  {\bibfnamefont {A.}~\bibnamefont {Siddharth}}, \bibinfo {author}
  {\bibfnamefont {Y.}~\bibnamefont {Popoff}}, \bibinfo {author} {\bibfnamefont
  {U.}~\bibnamefont {Drechsler}}, \bibinfo {author} {\bibfnamefont
  {D.}~\bibnamefont {Caimi}}, \bibinfo {author} {\bibfnamefont
  {S.}~\bibnamefont {H{\"o}nl}}, \bibinfo {author} {\bibfnamefont
  {J.}~\bibnamefont {Riemensberger}}, \bibinfo {author} {\bibfnamefont
  {J.}~\bibnamefont {Liu}}, \bibinfo {author} {\bibfnamefont {P.}~\bibnamefont
  {Seidler}},\ and\ \bibinfo {author} {\bibfnamefont {T.~J.}\ \bibnamefont
  {Kippenberg}},\ }\bibfield  {title} {\bibinfo {title} {A heterogeneously
  integrated lithium niobate-on-silicon nitride photonic platform},\ }\href
  {https://doi.org/10.1038/s41467-023-39047-7} {\bibfield  {journal} {\bibinfo
  {journal} {Nature Communications}\ }\textbf {\bibinfo {volume} {14}},\
  \bibinfo {pages} {3499} (\bibinfo {year} {2023})}\BibitemShut {NoStop}%
\end{thebibliography}
\end{document}